\documentclass[11pt]{article} 

\makeatletter\renewcommand{\section}{\@startsection
{section}{1}{\z@}{-3.5ex plus -1ex minus
    -.2ex}{2.3ex plus .2ex}{\bf }}

\makeatletter\renewcommand{\subsection}{\@startsection{subsection}{2}{\z@}{-3.25ex
plus -1ex minus
   -.2ex}{1.5ex plus .2ex}{\it }}
\makeatletter\renewcommand{\subsubsection}{\@startsection{subsubsection}{3}{-2.45ex}{-3.25ex
plus -1ex minus -.2ex}{1.5ex plus .2ex}{\it }}
\renewcommand{\thesection}{\arabic{section}}

\renewcommand{\theequation}{\thesection.\arabic{equation}}
\makeatletter \@addtoreset{equation}{section}

\usepackage{graphicx}
\usepackage{epsfig} 

\renewenvironment{thebibliography}[1]
     {\baselineskip=16pt plus 2pt minus 1pt
      \section*{\large\refname
        \@mkboth{\MakeUppercase\refname}{\MakeUppercase\refname}}%
     \list{\@biblabel{\@arabic\c@enumiv}}%
           {\settowidth\labelwidth{\@biblabel{#1}}%
            \leftmargin\labelwidth
            \advance\leftmargin\labelsep
            \@openbib@code
            \usecounter{enumiv}%
            \let\p@enumiv\@empty
            \renewcommand\theenumiv{\@arabic\c@enumiv}}%
      \sloppy
      \clubpenalty4000
      \@clubpenalty \clubpenalty
      \widowpenalty4000%
      \sfcode`\.\@m}

\let\fn\footnote
\renewcommand{\footnote}[1]{\linespread{1.1}\fn{#1}\linespread{1.29}}

\usepackage{amsfonts}
\usepackage{amssymb}
\usepackage{mathrsfs}
\usepackage{amsmath,amssymb}
\usepackage{bm}
\usepackage{caption}
\newcommand{\appendices}{\section*{Appendix}\setcounter{section}{0} \setcounter{equation}{0}
\renewcommand{\thesection}{\Alph{section}.}
\renewcommand{\theequation}{\thesection\arabic{equation}}}

\def\tyng(#1){\hbox{\tiny$\yng(#1)$}}

\newcommand{\be}{\begin{equation}}
\newcommand{\ee}{\end{equation}}
\newcommand{\bea}{\begin{array}}
\newcommand{\ea}{\end{array}}
\newcommand{\beqa}{\begin{eqnarray}}
\newcommand{\eeqa}{\end{eqnarray}}
\newcommand{\nn}{\nonumber}

\begin{document}
\begin{titlepage}
\begin{flushright}
\end{flushright}

\vskip 2 em

\begin{center}
\centerline{{\Large \bf A $U(3)$ Gauge Theory on Fuzzy Extra Dimensions}} 


\vskip 5em

\centerline{\large \bf S. K\"{u}rk\c{c}\"{u}o\v{g}lu and G. \"{U}nal}

\vskip 1em

\centerline{\sl $^\dagger$ Middle East Technical University, Department of Physics,}
\centerline{\sl Dumlupinar Boulevard, 06800, Ankara, Turkey}

\vskip 1em

{\sl kseckin@metu.edu.tr\,,} {\sl ugonul@metu.edu.tr \,} 

\end{center}
                                                                                                                                                                                                                                                                                                                                                                                                                                                                                                                                                                                                                                                                                                                                                                                                                                                                                                                                    
\vskip 4 em

\begin{quote}
\begin{center}
{\bf Abstract}
\end{center}

\vskip 1em

In this article, we explore the low energy structure of a $U(3)$ gauge theory over spaces with fuzzy sphere(s) as extra dimensions. In particular, we determine the equivariant parametrization of the gauge fields, which transform either invariantly or as vectors under the combined action of $SU(2)$ rotations of the fuzzy spheres and those $U(3)$ gauge transformations generated by $SU(2) \subset U(3)$ carrying the spin $1$ irreducible representation
of $SU(2)$. The cases of a single fuzzy sphere $S_F^2$ and a particular direct sum of concentric fuzzy spheres, $S_F^{2 \, Int}$, covering the monopole bundle sectors with windings $\pm 1$ are treated in full and the low energy degrees of freedom for the gauge fields are obtained. Employing the parametrizations of the fields in the former case, we determine a low energy action by tracing over the fuzzy sphere and show that the emerging model is abelian Higgs type with $U(1) \times U(1)$ gauge symmetry and possess vortex solutions on ${\mathbb R}^2$, which we discuss in some detail. Generalization of our formulation to the equivariant parametrization of gauge fields in $U(n)$ theories is also briefly addressed.

\vskip 1em

\vskip 5pt

\end{quote}

\end{titlepage}

\setcounter{footnote}{0}

\newpage

\section{Introduction}

It is by now very well-known that $N=4$ supersymmetric $SU({\cal N})$ Yang-Mills theories (SYM), deformed by the addition of cubic (soft supersymmetry breaking (SSB)) and mass terms in the scalar matter fields and relatedly $SU({\cal N})$ gauge theories coupled to a triplet of scalars carrying the adjoint representation of $SU({\cal N})$ as well as pure Yang-Mills (YM) matrix models with cubic and quadratic deformation terms develop fuzzy vacua, which are generically described by direct sums of products of fuzzy spheres ${\cal S}^2_F \times {\cal S}^2_F$ ($:= \oplus S_F^2 \times S_F^2$) or that of fuzzy spheres ${\cal S}_F^2 (:= \oplus S_F^2)$ \cite{Chatzistavrakidis:2009ix, Dorey:1999sj,Dorey:2000fc,Auzzi:2008ep, Aschieri:2006uw, Austing:2001ib, Ydri:2016dmy, Seckin-Derek, seckin-PRD2, seckin-PRD, Seckinson,Gonul-Seckin}. Such fuzzy sphere vacua also appear in BMN matrix models, which was proposed some time ago to give a non-perturbative description of the M-theory on maximally supersymmetric pp-wave backgrounds \cite{Berenstein:2002jq, Dasgupta:2002hx}. This is in fact a massive deformation of the BFSS matrix model \cite{Banks:1996vh}, while the latter was put forward to describe M-theory on flat backgrounds. Both $N=4$ SYM and BFSS emerge from dimensional reduction of $N=1$ SYM in $9+1$ dimensions, down to $3+1$ for SYM and to $0+1$ for the BFSS model \cite{Banks:1996vh, Erdmenger}, hence, it is rather not all too surprising to come across fuzzy vacuum configurations the deformations of one, once it is encountered in the other, albeit as ${\cal S}^2_F \times {\cal S}^2_F$ or ${\cal S}_F^2$ or vice versa. An alternative or a complementary perspective is gained by the fact that the BMN model can also be obtained from the $SU(2)_L$ -invariant dimensional reduction of $N=4$ SYM on ${\mathbb R} \times S^3$ \cite{Plefka}, where $S^3$ has the $SO(4) \approx SU(2)_L \times SU(2)_R$, symmetry.   

For the $SU(\cal N)$ YM theory on Minkowski space ${\mathbb M}^4$ coupled to a triplet of adjoint scalar fields fuzzy sphere $S_F^2$ vacuum was investigated in \cite{Aschieri:2006uw}. In this model, three matrices describing the $S_F^2$ are the vacuum expectation values(VEVs) of the scalars fields and the $SU(2)$ symmetry of $S_F^2$ is inherited from a global $SU(2)$ gauge symmetry of the YM model. Nonzero VEVs of the scalar fields also imply that the $SU(\cal N)$ gauge symmetry is spontaneously broken down to a $U(n)$, where ${\cal N}$, $n$ and the level $\ell$ of the fuzzy sphere are related as ${\cal N}= (2 \ell+1) n$. Fluctuations around this vacuum configuration are found to have the structure of $U(n)$ gauge fields over $S_F^2$, which preliminarily indicates that the emerging model after symmetry breaking may be conjectured to be an effective gauge theory over ${\mathbb M}^4 \times S_F^2$. A  Kaluza-Klein (KK) type mode expansion of the gauge fields and a detailed analysis of its low lying modes is performed in \cite{Aschieri:2006uw}, and places the effective gauge theory interpretation on firm grounds. 

Adaptation of coset space dimensional reduction (CSDR) techniques \cite{Forgacs, Zoupanos} (See also, \cite{Aschieri:2003vy} in this context) and equivariant parametrization of gauge fields into the framework of these models endows us with a complementary viewpoint in developing the effective gauge theory interpretation and understanding the low energy limit in this and a range of other models, which we have been recently investigating in \cite{Seckin-Derek, seckin-PRD2, seckin-PRD, Seckinson, Gonul-Seckin}. Equivariant parametrization method involves  imposing proper symmetry conditions on the fields of the model so that they transform covariantly under the action of the symmetry group of the extra dimensions up to the gauge transformations of the emergent model, i.e. those given by the gauge group surviving the symmetry breaking.  These conditions may be solved using the representation theory of Lie groups and explicit equivariant parametrizations of all the fields in the model can be obtained providing strong evidence for the interpretation of such models as effective gauge theories, since, subsequently, an effective low energy action (LEA) may be obtained by integrating out (i.e. tracing over) the fuzzy extra dimensions. Models with minimal non-Abelian gauge symmetry, $U(2)$ for the case of ${\mathcal M} \times {\cal S}_F^2$, and $U(4)$ for ${\mathcal M} \times {\cal S}^2_F \times {\cal S}^2_F$, where ${\mathcal M}$ denotes a Riemannian or a Lorentzian manifold, have been investigated in \cite{Seckin-Derek, seckin-PRD} and LEA were obtained when the extra dimensions do not have the direct sum structure but given by a single fuzzy sphere $S_F^2$ or $S_F^2 \times S_F^2$, respectively. LEA obtained in this manner, leads to Abelian Higgs type models\footnote{Some recent results obtained in the context of Aharony-Bergman-Jafferis-Maldacena (ABJM) models \cite{Aharony:2008ug, Nastase:2009ny}, has similarities with those of ours in \cite{Seckin-Derek, seckin-PRD, Seckinson, Gonul-Seckin}. 
 ABJM models are $N=6$ SUSY, $U({\cal N}) \times U({\cal N})$ Chern-Simons gauge theories at the level $(k, -k)$ which are coupled to scalar and spinor fields transforming respectively in the bifundamental and fundamental representation of its $SU(4)$ $R$-symmetry. A massive deformation of this model, which preserves the $N=6$ SUSY, but breaks the $R$-symmetry to $SU(2) \times SU(2) \times U(1)_A \times U(1)_B \times {\mathbb Z}_2$ was formulated in \cite{Gomis:2008vc,Terashima:2008sy} and it also has vacuum solutions which are fuzzy sphere(s). A certain parametrization for the fields in the bosonic sector of this model has been suggested in \cite{Mohammed:2012gi,Mohammed:2012rd}, which was shown to yield a low energy model in which four complex scalar fields interact with a sextic potential.} with vortex solutions for ${\mathcal M} \equiv {\mathbb R}^2$. There are several articles on the application of equivariant dimensional reduction method on higher dimensional YM gauge theories as well, for these \cite{Popov-Szabo, Lechtenfeld1, Popov, Popov2, Dolan-Szabo, Landi-Szabo,Lechtenfeld:2003cq} may be consulted. Other recent interesting articles within this general setting that we do not want to pass without mention include \cite{Zoupanos-1, Shahin1, Shahin2, Orfanidis,Steinacker:2014lma}. In \cite{Zoupanos-1}, for instance, an orbifold projection of $N=4$ SYM theory have been introduced and extra dimensions which are twisted fuzzy spheres consistent with this orbifolding were found to emerge due to the presence of SSB terms in the model. Models involving matrix valued fields in the adjoint of $SU(\cal N)$ have been proposed for inflation models in \cite{Shahin1,Shahin2}. Recently, new $4$- and $6$-dimensional fuzzy vacuum configurations in SSB deformed $N = 4$ SYM are reported in \cite{Steinacker:2014lma,  Steinacker:2014eua, Steinacker:2015mia}.

The outlined developments call for further investigations on the low energy structure around such fuzzy vacua in a diverse class of models with larger gauge groups in order to better assess the potential value of these models from a phenomenological point of view. In this article, we take a step forward and determine in full detail the equivariant field modes of a $U(3)$ gauge theory over ${\mathcal M} \times S^2_F$ and obtain the corresponding LEA by tracing over the fuzzy sphere. Firstly, we find that equivariant scalars, i.e. those fields transforming invariantly under the combined action of rotations and gauge transformations, may be constructed by taking advantage of the dipole and quadrupole terms, which appear in the branching of the adjoint representation of $SU(3)$ as $\underline{8} \rightarrow \underline{5} \oplus \underline{3}$ when the $SU(2)$ subgroup is maximally embedded in $SU(3)$. More concretely, we use these considerations and other group theoretical input coming from the equivariance conditions to construct the invariants as ``idempotents" involving intertwiners combining spin $\ell$ IRR of $SU(2)$ generating the the rotations of $S_F^2$ and those $U(3)$ gauge transformations generated by $SU(2) \subset U(3)$ carrying the spin $1$ IRR of $SU(2)$. There is also another invariant proportional to the ${\cal N}$-dimensional identity matrix, which essentially appears due to a $U(1)$ subgroup of $U(3) \approx SU(3) \times U(1)$. Equivariant vectors, i.e. those fields transforming as vectors under the equivariance constraints, are built using these invariants and the generators of $S_F^2$.  These developments are presented in section $3$ of our article, where we also show that the equivariance conditions break the $U(3)$ gauge symmetry down to the abelian product group $U(1) \times U(1) \times U(1)$. In section $4$ we obtain the LEA with this gauge symmetry, which, in addition to the three abelian gauge fields that naturally appear, contains two complex scalars each coupling to only one of the gauge fields and three real scalars interacting with the complex fields and with each other through a quartic potential. In the $\ell \rightarrow \infty $ limit, we determine the vacuum configuration of this quadric potential and use it in section $5$ to determine vortex solutions to the LEA,  in two different limits governed together by $\ell$ and the coupling constant of the constraint term in the potential, both of which is characterized by two winding numbers. Scattered through sections $3$ to $5$, we indicate how the commutative limit of our results relate to the instanton solutions in self-dual $SU(3)$ Yang-Mills theory for cylindrically symmetric gauge fields of Bais and Weldon \cite{Bais}. In particular, we point out the connection between the BPS vortices that we obtain in a certain commutative limit in section $5$ and the instanton solution in \cite{Bais}. In section $6$, we briefly outline the generalization of equivariant parametrization of gauge fields to $U(n)$ theories over ${\mathcal M} \times S^2_F$, and show that equivariant scalar are obtained by employing the $n-1$ multipole terms, that appear in the branching of the adjoint representation of $SU(n)$ under $SU(2)$, when the latter is maximally embedded in $SU(n)$. 

Section $7$ is devoted to the study of $U(3)$-equivariant fields over ${\mathcal M} \times S_F^{2 \, Int}$, where $S_F^{2 \, Int} := S_F^2 (\ell) \oplus S_F^2 (\ell) \oplus  S_F^2 \left ( \ell + \frac{1}{2} \right ) \oplus S_F^2 \left (\ell - \frac{1}{2} \right)$ was revealed in \cite{Seckinson} via a certain field redefinition of the triplet of scalars as a potentially interesting vacuum configuration for the $SU(\cal N)$ YM theory. The reason of interest on this vacuum is two fold. Firstly, through its certain projections it gives us access to fuzzy monopole bundles with winding numbers $\pm 1$, in a setting which is readily amenable to explicitly express the equivariant field modes and secondly it naturally identifies with the bosonic part of the $N=2$ fuzzy supersphere with $OSP(2,2)$ supersymmetry as discussed in \cite{Seckinson}. Let us note in passing that, we have in fact revealed a family of fuzzy vacua by generalizing the reparametrization performed over the scalar fields and developed these ideas further in a follow up work that focuses on the SSB and massive deformations of $N=4$ SYM, where we have also determined a family of fuzzy vacua given in terms of particular direct sums of products of fuzzy spheres and provided a detailed analysis of the equivariant fields in various sectors of an effective $U(4)$ gauge theory \cite{Gonul-Seckin}. In the present article, we are able to express all the equivariant field modes characterizing the low energy behaviour of the effective $U(3)$ theory on ${\mathcal M} \times S_F^{2 \, Int}$ in terms of suitable ``idempotents" and projection operators. From a geometrical point of view $S_F^{2 \, Int}$ vaccum is akin to stacks of concentric fuzzy D-branes carrying magnetic monopole fluxes, despite the fact that not all the string theoretic aspects \cite{Blumenhagen:2006ci} may be reproduced within the current framework \cite{Chatzistavrakidis:2009ix}. Nevertheless, this viewpoint allows us to think of the equivaraint gauge field modes of the effective gauge theory as those living on the world-volume of these D-branes, which may prove to be useful in an attempt to relate the effective gauge theory and the string theoretic perspectives. Let us finally note that we encounter in our analysis equivariant spinorial modes purely from group theoretical considerations, as has been already recognized in \cite{Seckinson, Gonul-Seckin}. Explicit expressions of these modes depend, in addition to the  ``idempotents" and projection operators, also on the vacuum value of an $SU(2)$ doublet which we introduce for the purpose of facilitating the aforementioned field redefinition. Evidently, these spinorial modes do not constitute independent degrees of freedom as they are introduced through a field redefinition, however, they play the role of the ``square roots" of the independent equivariant gauge field modes as the latter may be constructed by taking their suitable bilinears. 

\section{$U(n)$ Gauge Theory over ${\cal M}\times S_F^{2}$}

In order to orient the developments, we start with briefly explaining how an $SU(\cal N)$ gauge theory coupled to a triplet of adjoint scalar fields develop extra dimensions in the form of a fuzzy sphere $S_F^2$ \cite{Aschieri:2006uw}. We are interested in the model whose action may be given as
\begin{align}
S&=\int_{\cal M}d^dy\,Tr_{\cal N}\bigg (- \frac{1}{4g^2} F^{\dagger}_{\mu \nu} F^{\mu\nu} - (D_{\mu}\Phi_a)^\dagger (D^{\mu}\Phi_a)\bigg)- \frac{1}{\tilde{g}^2}V_1(\Phi) - a^2 V_2(\Phi) \,,\label{action}
\\\nonumber
\\& V_1(\Phi) = Tr_{\cal N}(F^{\dagger}_{ab}F_{ab}),\quad V_2(\Phi)=Tr_{\cal N}((\Phi_a\Phi_a+\tilde{b}\bm{1}_{\cal N})^2) \,,
\label{potterms}
\end{align} 
where $g,\tilde g,a, \tilde b$ are constants and $Tr_{\cal N}={\cal N}^{-1}Tr$ indicate a normalized trace. Here $\Phi_a\,(a=1,2,3)$ are anti-Hermitian scalar fields transforming in the adjoint representation of $SU(\cal N)$ as
\begin{align}
 \Phi_a\rightarrow U^\dagger \Phi_aU\,,\quad U\in SU(\cal N)\,,
 \label{gt}
\end{align}
and $A_\mu$ are $su(\cal N)$ valued anti-Hermitian gauge fields associated to $F_{\mu\nu}$. In the potential term $V_1(\Phi)$, $F_{ab}$ are defined as 
\begin{align}
 F_{ab}:=[\Phi_a,\Phi_b]-\epsilon_{abc}\Phi_c\,.
\end{align}
$V_2(\Phi)$ is a constraint term, whose purpose is essentially to force the model to select the single fuzzy sphere $S_F^2$ vacuum configuration, as opposed to a vacuum given in terms of the direct sums of fuzzy spheres, say, : ${\cal S}_F^2 := \oplus S_F^2$. We may also note that the Manifold ${\cal M}$, may be selected as a $d$-dimensional manifold on which \eqref{action} is renormalizable. In particular, it may be taken as the $4$-dimensional Minkowski space or ${\mathbb R}^2$ as we do so in section $5$.

It is obvious that potential terms $V_1(\Phi)$ and $V_2(\Phi)$ are positive definite and the minimum of potentials can be obtained by solving the equations
\begin{align}
F_{ab}=[\Phi_a,\Phi_b]-\epsilon_{abc}\Phi_c=0 \,,\quad -\Phi_a\Phi_a=\tilde{b}\bm{1}_{\cal N}\,.
\label{minpot}
\end{align}
A well-known solution \cite{Aschieri:2006uw} to these equations is given by taking $\tilde b$ as the eigenvalue of the quadratic Casimir of an irreducible representation (IRR) $\ell$ of $SU(2)$, and assuming that the dimension $\cal N$ of the matrices $\Phi_a$ factorize as ${\cal N} = (2\ell+1)n$. Then, up to the gauge transformations (\ref{gt}) the matrices
\begin{align}
\Phi_a=X_a^{(2\ell+1)}\otimes \bm{1}_n\,,
\label{vacuum}
\end{align}
where $X_a^{(2\ell+1)}$ are the anti-Hermitian generators of $SU(2)$ in the irreducible representation $\ell$ with the commutation relation
\begin{align}
[X_a^{(2\ell+1)},X_b^{(2\ell+1)}] = \epsilon_{abc}X_c^{(2\ell+1)}\,.
\end{align} 
satisfy (\ref{minpot}).

Evidently, this vacuum configuration breaks $SU(\cal N)$ symmetry down to $U(n)$. In addition, we see that it may be interpreted as the fuzzy sphere at level $\ell$ since the latter, at level $\ell$, is the algebra of $(2\ell+1)\times(2\ell+1)$ matrices generated by the three Hermitian coordinate functions
\begin{align}
 \hat{x}_a:=\frac{i}{\sqrt{\ell(\ell+1)}}X_a^{(2\ell+1)}
\end{align}
satisfying
\begin{align}
[\hat{x}_a,\hat{x}_b]=\frac{i}{\sqrt{\ell(\ell+1)}}\epsilon_{abc}\hat{x}_c\,,\quad \hat{x}_a\hat{x}_a=1\,.
\end{align} Derivatives on $S_F^2(\ell)$ are given by the derivations on the matrix algebra, which are simply implemented by the adjoint action of $su(2)$ on $S_F^2$
\begin{align}
 f\rightarrow ad X_a^{(2\ell+1)} f:=[X_a^{(2\ell+1)},f]\,,\quad f\in Mat(2\ell+1)\,.
\end{align}
In the commutative limit $\ell \rightarrow \infty$, $\hat{x}_a$ converge to the standard coordinates $x_a$ on $\mathbb{R}^3$, restricted to the unit sphere $x_ax_a=1$ and the derivations $[X_a^{(2\ell+1)},\cdot]$ become the vector fields $-i\mathcal{L}_a=\epsilon_{abc}x_b\partial_c$ .

Fluctuations about the vacuum (\ref{vacuum}) are given as 
\begin{align}
 \Phi_a=X_a+A_a\,,
\end{align}
where the short-hand notation $X_a^{(2\ell+1)}\otimes \bm{1}_n=X_a$ has been introduced. A short calculation yields that 
\begin{align}
F_{ab}=[X_a,A_b] - [X_b,A_a]+[A_a,A_b]-\epsilon_{abc}A_c \,,
\end{align}
which has the form of a curvature tensor for $U(n)$ gauge fields over $S_F^2$. This suggests that, the model emerging after spontaneous symmetry breaking as a $U(n)$ gauge theory on ${\cal M}\times S_F^2$ with the gauge fields $A_M(y)=(A_\mu(y),A_a(y))\in u(n)\otimes u(2\ell+1)$ and the field strength tensor $F_{MN}=(F_{\mu\nu},F_{a\mu},F_{ab})$
\begin{align}
F_{\mu\nu}&=\partial_\mu A_\nu-\partial_\nu A_\mu+[A_\mu,A_\nu]\,,\nonumber
\\F_{\mu a}&=D_\mu\Phi_a=\partial_\mu\Phi_a+[A_\mu,\Phi_a]=\partial_\mu A_a-[X_a,A_\mu]+[A_\mu,A_a]\,,
\\F_{ab}&=[\Phi_a,\Phi_b]-\epsilon_{abc}\Phi_c=[X_a,A_b]-[X_b,A_a]+[A_a,A_b]-\epsilon_{abc}A_c\,.\nonumber
\end{align} 

It is well-known fact that on fuzzy sphere there are three components of the gauge field $A_a$, which can only be disentangled from each other in the commutative limit. On $S^2$, there are only two degrees of freedom for the gauge field $A_a$ and the standard treatment is to impose the constraint $x_aA_a=0$ to eliminate the normal component of $A_a$. Here the constraint term $V_2$ in (\ref{potterms}) serves the purpose of suppressing the normal component of $A_a$ by giving it a large mass $a\sqrt{\ell(\ell+1)}$, as $\ell\rightarrow \infty$, \cite{Seckin-Derek, Aschieri:2006uw}. In the discussion above we have worked with dimensionless $\Phi_a$. We can restore the dimensions by taking $\Phi_a \rightarrow  \gamma \Phi_a$ where $\gamma$ has the mass dimensions $[m]^{d/2-1}$. Working with the dimensionful $\Phi_a$'s, we have the mass dimension of the couplings $g$ and $\tilde g$ are $[g]=[m]^{-d/2+2}$ and $[\tilde{g}]=[m]^{d/2-2}$. We also note that performing the scaling ${\tilde \Phi}_a =  \sqrt{2} g \Phi_a$ and taking $g\tilde{g} = 1$, the part of the action without the constraint term, $V_2(\Phi)$, may be expressed as the $L^2$-norm of $F_{MN}$ and we may write
\begin{align}
S= \frac{1}{4 g^2} \int d^d\,y \, Tr_{n(2 \ell +1)} F_{MN}^\dagger F^{MN}+V_2(\Phi)\,.
\label{stym}
\end{align}

A Kaluza-Klein mode expansion of the gauge fields over fuzzy extra dimension given in \cite{Aschieri:2006uw}, and an inspection of its low lying modes supports the effective gauge theory interpretation. A complementary approach in the context is the equivariant parametrization technique which entails imposing proper symmetry conditions on the fields of the model so that they transform covariantly under the action of the symmetry group of the extra dimensions up to gauge transformations of the emergent model. As discussed in the introduction, we now take up the task of examining the $U(3)$ model on ${\cal M}\times S_F^2$ by employing this method.

\section{$SU(2)$-equivariant gauge fields for $U(3)$ gauge theory}

Here, our initial aim is to construct the explicit form of $SU(2)$-equivariant gauge fields in this $U(3)$ theory. To be somewhat more precise, we will determine those field configurations which are transforming as scalars and vectors under rotations of $S_F^2$ up to $U(3)$ gauge transformation. For this purpose, we introduce the infinitesimal symmetry generators $\omega_a$ as
\begin{align}
\omega_a=X_a^{(2\ell+1)}\otimes \bm{1}_3-\bm{1}_{(2\ell+1)}\otimes i\Sigma_a \,,
\end{align} 
where $\Sigma_a$ are the spin $1$ irreducible representation of $SU(2) \subset SU(3)$: ${(\Sigma_a)}_{ij}=i\epsilon_{iaj}$ and $\omega_a$ satisfy the condition
\begin{align}
[\omega_a,\omega_b]=\epsilon_{abc}\omega_c\,.
 \label{omega}
\end{align}
Clearly, the adjoint action $ad\, \omega_a \cdot=[\omega_a,\cdot]$, is composed of infinitesimal rotations over $S_F^2$ combined with those infinitesimal $SU(3)$ transformations, which are generated by $\Sigma_a$.

In fact, the adjoint representation of $SU(3)$ decomposes to $SU(2)$ IRR's as
\begin{align}
\underline{8} \rightarrow \underline{5} \oplus \underline{3} \,.
\label{su3}
\end{align}
In this branching, $\Sigma_a$ generate the $\underline{3}$ (spin $1$) IRR of $SU(2)$, while the remaining five generators of $SU(3)$ may be given in the form of the quadrupole tensor
\begin{align}
 &Q_{ab}=\frac{1}{2}\lbrace\Sigma_a,\Sigma_b\rbrace-\frac{2}{3}\delta_{ab}\,,
 \\&{(Q_{ab})}_{ij}=\delta_{ai}\delta_{bj}+\delta_{aj}\delta_{bi}-\frac{2}{3}\delta_{ab}\delta_{ij} \,,
\end{align} 
carrying the spin $2$ (i.e $\underline{5}$) IRR of $SU(2)$. For each IRR of $SU(2)$ in the branching (\ref{su3}), we may expect to construct one rotational invariant under $ad\, \omega_a$ in addition to the identity matrix $\bm{1}_{(2\ell+1)3}$ and we will at once proceed to see that this is indeed so\footnote{Generalization of this construction to all $U(n)$ gauge theories on ${\mathcal M} \times S_F^2$ is discussed in section \ref{Generalization}.}. These invariants may be simply taken as $X_a\Sigma_a$ and $X_aX_b Q_{ab}$, however we prefer to express them as``idempotent" matrices, which turn out to be suitable for the subsequent construction of the equivariant vectors, as well as for clarity.

In order to find the $SU(2)$-equivariant gauge fields, we impose the following symmetry constraints
\be
[\omega_a, A_\mu]=0\,, \quad [\omega_a,A_b]=\epsilon_{abc}A_c\,,
\label{symcons}
\ee 
which simply imply that, under the adjoint action of $\omega_a$, $A_\mu$ are rotational invariants and $A_a$ transform as vectors.

$SU(2)$ IRR content of $\omega_a$ may be found by the following tensor product
\begin{align}
\ell\otimes 1=(\ell-1)\oplus \ell\oplus (\ell+1)\,,
\end{align} 
and therefore IRR decomposition of the adjoint action of $\omega_a$ is
\begin{align}
[(\ell-1)\oplus \ell\oplus (\ell+1)]\otimes [(\ell-1)\oplus \ell\oplus (\ell+1)]= {\bm 3} 0 \oplus {\bm 7} 1 \oplus \cdots\,.\label{Clebsch}
\end{align} 
where the coefficients in bold denote the multiplicities of respective IRR in front of which they appear. From this Clebsch-Gordan expansion, it can be seen that the set of solutions for $A_\mu$ is $3$-dimensional. We span this space by the invariants $Q_1,Q_2$, as defined below and $\mathbf{1}_{(2\ell+1)3}$ and introduce the following explicit parametrization of $A_\mu$: 
\begin{align}
A_\mu=-\frac{1}{2}a_\mu^{(1)}(y)Q_1+\frac{1}{2} a_\mu^{(2)}(y)Q_2+\frac{i}{2}\bigg(\frac{a_\mu^{(1)}(y)-a_\mu^{(2)}(y)}{3}+b_\mu(y)\bigg)\bm{1}\,,
\label{scalars}
\end{align} 
where $a_\mu^{(1)},a_\mu^{(2)},b_\mu$ are Hermitian $U(1)$ gauge fields\footnote{The reason for this particular form of the coefficients of $Q_1\,,Q_2$ and $\bm{1}$ in (\ref{scalars}) will become clear as we proceed to perform the dimensional reduction over $S_F^2$ in the next section.} on $\cal M$ and $Q_1,Q_2$ are anti-Hermitian idempotents given as \cite{Bal}
\begin{align}
Q_1&=\frac{2(iX_a\Sigma_a+\ell+1)(iX_b \Sigma_b+1)-(\ell+1)(2\ell+1)\mathbf{1}}{i(\ell+1)(2\ell+1)}\,,\quad Q_1^\dagger=-Q_1\,,\quad Q_1^2=-\bm{1}_{3(2\ell+1)}\,,\nonumber
\\Q_2&=\frac{2(iX_a\Sigma_a-\ell)(iX_b\Sigma_b+1)-\ell(2\ell+1)\mathbf{1}}{i\ell(2\ell+1)}\,,\quad Q_2^\dagger=-Q_2\,,\quad Q_2^2=-\bm{1}_{3(2\ell+1)}\,.\label{idem}
\end{align}
Thus, we see that $U(3)$ gauge symmetry is broken down to $U(1) \times U(1) \times U(1)$. Under the gauge transformation generated by $U = e^{ -\frac{1}{2} \theta_1(y) Q_1}e^{ \frac{1}{2} \theta_2(y) Q_2}e^{i \left ( \frac{1}{6} \theta_1(y) -  \frac{1}{6} \theta_2(y) + \frac{1}{2} \theta_3(y) \right ) {\bm 1}}$, it is readily seen that $A_\mu \rightarrow A_\mu^\prime$ with $a_\mu^{(i) \prime} = a_\mu^{(i)} +\partial_\mu \theta_i$ and $b_\mu^{\prime} = b_\mu + \partial_\mu \theta_3$, hence the rotationally symmetry of $A_\mu$ is preserved.  

Equation (\ref{Clebsch}) shows that the dimension of the set of solutions for $A_a$ is seven and its parametrization may be chosen as follows
\begin{multline}
 A_a=\frac{1}{2}\varphi_1(y)[X_a,Q_1]+\frac{1}{2}\chi_1(y)[X_a,Q_2]-\frac{1}{2}(\varphi_2(y)+1)Q_1[X_a,Q_1]+\frac{1}{2}(\chi_2(y)-1)Q_2[X_a,Q_2]
\\ +\frac{i}{2}\frac{\varphi_3(y)}{2(\ell+1/2)}\bigg(\lbrace X_a,Q_1\rbrace-iQ_2[X_a,Q_2]\bigg)+\frac{i}{2}\frac{\chi_3(y)}{2(\ell+1/2)}\bigg(\lbrace X_a,Q_2\rbrace-iQ_1[X_a,Q_1]\bigg)
 \\+\frac{1}{2}\psi(y)\frac{\omega_a}{\ell+1/2}\,.
 \label{vectors}
\end{multline} 
Let us digress for a moment and inspect (\ref{vectors}) in some detail. Observe that we have essentially used commutators and anticommutators of $Q_1$ and $Q_2$ with $X_a$ to construct a suitable basis for vectors fulfilling (\ref{symcons}). As coefficients of these vectors, we have introduced the real scalar fields $\varphi_1\,,\varphi_2\,,\varphi_3\,,\chi_1\,,\chi_2\,,\chi_3$ and $\psi$ on $\cal M$. We will see shortly that some of these naturally combine to form complex scalars when the model is dimensionally reduced over $S_F^2$.

In the commutative limit, $\ell \rightarrow \infty$ ($\frac{X_a}{\ell}\rightarrow \hat{x}_a\,, \hat{x}_a\hat{x}_a=1$), we have 
\begin{align}
 &iQ_1=q_1=(\Sigma_a\hat{x}_a)^2+(\Sigma_a\hat{x}_a)-1\,,\nonumber
 \\&iQ_2=q_2=(\Sigma_a\hat{x}_a)^2-(\Sigma_a\hat{x}_a)-1\,,
 \label{commutativeidem}
\end{align} where $q_1^2=q_2^2=\bm{1}_3$. Another idempotent may be given as a linear combination of $q_1$ and $q_2$ and ${\bm 1}_3$ as $q_3=-(q_1+q_2)-\bm{1}_3$ \cite{Bal}. Using (\ref{commutativeidem}), we find that the commutative limit of $A_a$ in (\ref{vectors}) takes the form 
\begin{align}
 A_a\xrightarrow[\ell\rightarrow \infty]{}-\frac{\varphi_1(y)}{2}\mathcal{L}_a q_1-\frac{\chi_1(y)}{2}\mathcal{L}_aq_2-i\frac{(\varphi_2(y)+1)}{2}q_1\mathcal{L}_aq_1+i\frac{(\chi_2(y)-1)}{2}q_2\mathcal{L}_aq_2+\frac{\varphi_3(y)}{2}\hat{x}_aq_1\nonumber
 \\+\frac{\chi_3(y)}{2}\hat{x}_aq_2+\frac{\psi(y)}{2}\hat{x}_a\,.
 \label{commutativeA_a}
\end{align}
Imposing the constraint $x_aA_a=0$ eliminates the radial component of the gauge field. We see from (\ref{commutativeA_a}) that this condition is satisfied 
if and only if we set $\varphi_3=\chi_3=\psi=0$. The remaining terms of $A_a$ in (\ref{commutativeA_a}) and the commutative limit of $A_\mu$ (apart from a $b_\mu$-field due to the $U(1)$ subgroup of $U(3)$, which decouples from the rest in the commutative limit, as we shall explicitly see later on in the section \ref{vor}) are in agreement with the cylindrical symmetric ansatz for the $SU(3)$ Yang-Mills theory of Bais and Weldon \cite{Bais}. 

\section{Dimensional reduction of the Yang-Mills action}

In this section, we pursue the dimensional reduction of our model over $S_F^2$. We can substitute our equivariant gauge fields $A_\mu$ and $A_a$ into the action (\ref{action}), and then by tracing over the fuzzy sphere $S_F^2$, we obtain the reduced action on $\cal M$. The following identities are very useful to simplify the calculations 
\begin{align}
 &[X_a,\lbrace X_a,Q_i\rbrace]=0\,,\quad [Q_i,\lbrace X_a,Q_i\rbrace]=0\,,\quad \lbrace X_a,[X_a,Q_i]\rbrace=0\,,\quad \lbrace Q_i,[X_a,Q_i]\rbrace=0\,,
\end{align}
where $i=1,2$ and sum over only the repeated index ``a" is implied.

Borrowing the notation of \cite{Seckin-Derek},
\begin{align}
 S=\int_{\cal M}d^d y \, \, ({\cal L}_F+{\cal L}_G+\frac{1}{\tilde{g}^2}V_1+a^2V_2)\,.
 \label{notac}
\end{align}
Now, we start to calculate each term in (\ref{notac}) separately. For the field strength term, the curvature $F_{\mu\nu}$ can be expressed in terms of the rotational invariants $Q_1\,,Q_2$ and $\bm{1}$ as
\begin{align}
 F_{\mu\nu}= - \frac{1}{2} f^{(1)}_{\mu\nu}Q_1 + \frac{1}{2} f^{(2)}_{\mu\nu}Q_2 + i \frac{1}{2} \left (\frac{f^{(1)}_{\mu\nu}-f^{(2)}_{\mu\nu}}{3}+h_{\mu\nu} \right ) {\bm 1}
\end{align}
where we have introduced
\begin{align}
 f^{(1)}_{\mu\nu} := \partial_\mu a_\nu^{(1)}-\partial_\nu a_\mu^{(1)}\,,\quad f^{(2)}_{\mu\nu} := \partial_\mu a_\nu^{(2)}-\partial_\nu a_\mu^{(2)}\,,\quad h_{\mu\nu} := \partial_\mu b_\nu-\partial_\nu b_\mu\,.
\end{align}
Then, ${\cal L}_F$ takes the form
\begin{multline}
 {\cal L}_F:=\frac{1}{4g^2}Tr_{\cal N}(F_{\mu\nu}^\dagger F^{\mu\nu})
 \\=\frac{1}{g^2}\Bigg (\frac{\ell+1}{9(2\ell+1)} f^{(1)}_{\mu\nu} {f^{(1)}}^{\mu\nu}+\frac{\ell}{9(2\ell+1)}f^{(2)}_{\mu\nu} {f^{(2)}}^{\mu\nu}+\frac{1}{18}f^{(1)}_{\mu\nu}{f^{(2)}}^{\mu\nu}+\frac{1}{16}h_{\mu\nu}h^{\mu\nu}+\frac{1}{6(2\ell+1)}f^{(1)}_{\mu\nu}h^{\mu\nu}
 \\+\frac{1}{6(2\ell+1)}f^{(2)}_{\mu\nu}h^{\mu\nu} \Bigg)\,.
\end{multline}
The covariant derivative term $D_\mu\Phi_a$ is calculated to be
\begin{multline}
 D_\mu\Phi_a=\frac{1}{2}\left(D_\mu \varphi_1\right)[X_a,Q_1]+\frac{1}{2}\left(D_\mu\chi_1\right)[X_a,Q_2]-\frac{1}{2}\left(D_\mu\varphi_2\right)Q_1[X_a,Q_1]+\frac{1}{2}\left(D_\mu\chi_2\right)Q_2[X_a,Q_2]
 \\+\frac{i}{4}\frac{\partial_\mu\varphi_3}{(\ell+1/2)}(\lbrace X_a,Q_1\rbrace-iQ_2[X_a,Q_2])+\frac{i}{4}\frac{\partial_\mu\chi_3}{(\ell+1/2)}(\lbrace X_a,Q_2\rbrace-iQ_1[X_a,Q_1]) \\ 
 +\frac{1}{2(\ell+1/2)}\left(\partial_\mu \psi\right)\omega_a\,,
\end{multline}
where $D_\mu\varphi_i=\partial_\mu \varphi_i+\epsilon_{ji}a_\mu^{(1)}\varphi_j$ and $D_\mu\chi_i=\partial_\mu \chi_i+\epsilon_{ji}a_\mu^{(2)}\chi_j$. After tracing, the gradient term ${\cal L}_G$ reads 
\begin{align}
 {\cal L}_G&=Tr((D_\mu\Phi_a)^\dagger D_\mu\Phi_a)
 \\=&\frac{2\ell(2\ell+3)}{3(\ell+1)(2\ell+1)}((D_\mu\varphi_1)^2+(D_\mu\varphi_2)^2)+\frac{2(2\ell-1)(\ell+1)}{3\ell(2\ell+1)}((D_\mu\chi_1)^2+(D_\mu\chi_2)^2)\nonumber
 \\&+\frac{6\ell^5+15\ell^4+4\ell^3-9\ell^2+2}{3\ell(\ell+1)(2\ell+1)^3}((\partial_\mu\varphi_3)^2+(\partial_\mu\chi_3)^2)+\frac{\ell^2+\ell+2}{(2\ell+1)^2}(\partial_\mu\psi)^2\nonumber
 \\&-\frac{2\ell(\ell+1)}{3(2\ell+1)^2}\partial_\mu\varphi_3\partial_\mu\chi_3-\frac{2\ell(2\ell^2-5\ell-9)}{3(2\ell+1)^3}\partial_\mu\psi\partial_\mu\varphi_3-\frac{2(2\ell^3+11\ell^2+7\ell-2)}{3(2\ell+1)^3}\partial_\mu\chi_3\partial_\mu\psi\,.
\end{align}We note that $\varphi_1\,,\varphi_2$ and $\chi_1\,,\chi_2$ naturally combine to two complex scalar fields $\varphi:=\varphi_1+i\varphi_2\,,\chi:=\chi_1+i\chi_2$, with $D_\mu \varphi = (\partial_\mu + i a_\mu^{(1)}) \varphi$ and $D_\mu \chi = (\partial_\mu + i a_\mu^{(2)}) \chi$, which we will make use of in the next section.

In order to calculate the potential term $V_1$, it is useful to work with the dual of the curvature $F_{ab}$. We find
\begin{align}
 \frac{1}{2}\epsilon_{abc}F_{ab}=\Lambda_1+\Lambda_2|\varphi|^2+\Lambda_3|\chi|^2+\Lambda_4(\varphi_3^2+\chi_3^2)+\Lambda_5\varphi_3+\Lambda_6\chi_3+\Lambda_7\varphi_3\chi_3+\Lambda_8\varphi_3\psi\nonumber
 \\+\Lambda_9\chi_3\psi+\Lambda_{10}(\varphi_1+\varphi_2Q_1)[X_a,Q_1]+\Lambda_{11}(\chi_1+\chi_2Q_2)[X_a,Q_2]+\Lambda_{12}\psi+\Lambda_{13}\psi^2\,,
 \label{dualF}
\end{align} 
where $\Lambda_i\,,i=1,\cdots,11$ are the $3(2\ell+1)\times 3(2\ell+1)$ dimensional matrices which are listed in the appendix. Using (\ref{dualF}), the potential term $V_1$ may be determined as
\begin{align}
 V_1=Tr_{\cal N}(F_{ab}^\dagger F_{ab})&=\alpha_1-\alpha_2|\varphi|^2-\alpha_3|\chi|^2-\alpha_4\varphi_3^2-\alpha_5\chi_3^2-\alpha_6\varphi_3+\alpha_7\chi_3-\alpha_8\varphi_3\chi_3-\alpha_9\varphi_3\psi\nonumber
 \\&-\alpha_{10}\chi_3\psi+\alpha_{11}\psi^2+\beta_{1}|\varphi|^4-\beta_2|\varphi|^2|\chi|^2+\beta_{3}|\varphi|^2\varphi_3^2+\beta_{4}|\varphi|^2\chi_3^2-\beta_{5}|\varphi|^2\varphi_3\nonumber
 \\&+\beta_6|\varphi|^2\chi_3-\beta_7|\varphi|^2\varphi_3\chi_3+\beta_8|\varphi|^2\varphi_3\psi-\beta_9|\varphi|^2\chi_3\psi+\beta_{10}|\varphi|^2\psi^2+\gamma_1|\chi|^4\nonumber
 \\&-\gamma_2|\chi|^2\varphi_3^2+\gamma_3|\chi|^2\chi_3^2+\gamma_4|\chi|^2\varphi_3-\gamma_5|\chi|^2\chi_3+\gamma_6|\chi|^2\varphi_3\chi_3-\gamma_7|\chi|^2\varphi_3\psi\nonumber
 \\&-\gamma_8|\chi|^2\chi_3\psi+\gamma_9|\chi|^2\psi^2-\delta_1(\varphi_3^4+\chi_3^4+6\varphi_3^2\chi_3^2)-\delta_2(\varphi_3^3+3\varphi_3\chi_3^2)\nonumber
 \\&-\delta_3(\chi_3^3+3\chi_3\varphi_3^2)-\delta_4(\varphi_3^3\chi_3+\chi_3^3\varphi_3)-\delta_5(\varphi_3^3\psi+3\varphi_3\chi_3^2\psi)\nonumber
 \\&-\delta_6(\chi_3^3\psi+3\chi_3\varphi_3^2\psi)+\delta_7(\varphi_3^2\psi+\chi_3^2\psi)+\delta_8(\varphi_3^2\psi^2+\chi_3^2\psi^2)\nonumber
 \\&-\delta_{9}\varphi_3\chi_3\psi-\delta_{10}\varphi_3\psi^2-\delta_{11}\chi_3\psi^2-\delta_{12}\varphi_3\chi_3\psi^2-\delta_{13}\varphi_3\psi^2\nonumber
 \\&-\delta_{14}\chi_3\psi^3-\delta_{15}\psi^3-\delta_{16}\psi^4\,,
\label{p1}
\end{align} 
where all the $\ell$-dependent constants: $\alpha\,,\beta\,,\gamma\,,\delta$ are given in the appendix.

In the $\ell \longrightarrow \infty$ limit we find
\begin{multline}
V_1(\Phi) \Big |_{\ell \rightarrow \infty}
=  \frac{2}{3} (|\varphi|^2+\varphi_3 -1)^2 + \frac{2}{3} (|\chi|^2 - \chi_3 -1)^2 + \frac{2}{3} (|\varphi|^2-|\chi|^2)^2 + \frac{4}{3} |\varphi|^2\varphi_3^2 + \frac{4}{3} |\chi|^2\chi_3^2  \\
- \frac{1}{6}\varphi_3^2 - \frac{1}{6} \chi_3^2+ \frac{1}{2} \psi^2 - \frac{1}{3}(\varphi_3\chi_3+\varphi_3\psi+\chi_3\psi) \,.
\label{pinf1}
\end{multline}

The potential $V_1(\Phi)=Tr_{\cal N}(F_{ab}^\dagger F_{ab})$ is positive definite, although the r.h.s of (\ref{p1}) and (\ref{pinf1}) are not manifestly so. For the limiting case (\ref{pinf1}) we have determined that minima occurs at the following configurations
\beqa
i) \quad |\varphi|^2&=&0 \,, \quad |\chi|^2=1 \,, \quad \varphi_3=\chi_3=\psi=0 \,, 
\label{absminima} \\
ii) \quad |\varphi|^2&=&0 \,, \quad |\chi|^2=0 \,, \quad \varphi_3=1 \,, \chi_3=-1 \,, \psi = 0 \,, \\
iii) \quad |\varphi|^2&=&\frac{1}{\sqrt{2}} \,, \quad  |\chi|^2=0 \,, \quad \varphi_3= 0 \,, \quad  \chi_3 = -\frac{3}{2}  \,, \quad  \psi = - \frac{1}{2} \,, \\
iv) \quad |\varphi|^2&=&0 \,, \quad  |\chi|^2=\frac{1}{\sqrt{2}} \,, \quad \varphi_3= \frac{3}{2} \,, \quad \chi_3= 0 \,, \quad  \psi= \frac{1}{2} \,.
\eeqa

For the computation of the last term in (\ref{notac}), we first obtain the expression
\begin{align}
 \Phi_a\Phi_a+\ell(\ell+1)=R_1+R_2iQ_1+R_3iQ_2\,,
 \label{constraint}
\end{align} 
where $R_1\,,R_2$ and $R_3$ are listed in the appendix. Then, the potential term $V_2$ is determined to be
\begin{align}
V_2(\Phi) = \bigg(R_1^2+R_2^2+R_3^2-\frac{2(2\ell-3)}{3(2\ell+1)}R_1R_2-\frac{2(2\ell+5)}{3(2\ell+1)}R_1R_3-\frac{2}{3}R_2R_3\bigg)\,.
\end{align}
In the large $\ell$ limit we find
\beqa
a^2 V_2(\Phi) \Big |_{\ell \rightarrow \infty} &=& \frac{1}{3} a^2 \left ( (R_1-R_2-R_3)^2 + (-R_1+R_2-R_3)^2 +  (-R_1-R_2+R_3)^2 \right )\Big |_{\ell \rightarrow \infty} \,, \nn \\ 
&=&  \frac{1}{3} a^2 \ell^2 \left ((-\psi + \varphi_3 + \chi_3)^2 + (\psi - \varphi_3 + \chi_3)^2 +  (\psi + \varphi_3 - \chi_3)^2 \right) \,.
\label{minimav2}
\eeqa 
In the next section we will first consider the scaling limit $a \rightarrow 0$, $\ell \rightarrow \infty$, with $a \ell $ kept finite but small. Then, among the minima of the potential $V_1(\Phi)$ listed above, only (\ref{absminima}) minimizes (\ref{minimav2}) as can easily be observed.

\section{Vortices}\label{vor}

In this section, we would like to inspect the structure of the reduced action (\ref{notac}) on ${\mathcal M} \equiv {\mathbb R}^2$ and show that it has static vortex type solutions. We are interested in exploring these in two different limits, namely, $i)$ $\ell\rightarrow \infty\,,a \rightarrow 0$ with $a \ell $ remaining finite but small and $ii)$ $a\rightarrow \infty$ and $\ell$ is large but finite. These limits are physically well-motivated since in the absence of any canonical choices for the parameter $a$, they give the two extremes for handling the constraint term $V_2(\Phi)$.
 
\subsection{Case $i)$}

In this case the reduced action becomes
\begin{align}
 S=\int d^2 y \, \,\bigg(\frac{1}{18g^2}(f^{(1)}_{\mu\nu}{f^{(1)}}^{\mu\nu}+f^{(2)}_{\mu\nu}{f^{(2)}}^{\mu\nu}+f^{(1)}_{\mu\nu}{f^{(2)}}^{\mu\nu})+\frac{1}{16g^2}h_{\mu\nu}h^{\mu\nu}+\frac{2}{3}(|D_\mu\varphi|^2+|D_\mu\chi|^2)\nonumber
 \\+\frac{1}{4}((\partial_\mu\varphi_3)^2+(\partial_\mu\chi_3)^2+(\partial_\mu\psi)^2)-\frac{1}{6}(\partial_\mu\varphi_3\partial_\mu\chi_3+\partial_\mu\varphi_3\partial_\mu\psi+\partial_\mu\chi_3\partial_\mu\psi) +  \frac{1}{\tilde{g}^2} V_1(\Phi) \Big |_{\ell \rightarrow \infty} \bigg) \,. 
\label{action2}
\end{align} 
We observe that, the gauge field $b_\mu$ decouples from the rest of the action, and does not play any role in the rest of this subsection. Thus we essentially have a abelian Higgs type model with $U(1) \times U(1)$ gauge symmetry. The vacuum configuration is given by (\ref{absminima}) and has the structure of $T^2 = S^1 \times S^1$, with $\pi_1(T^2) = {\mathbb Z} \oplus {\mathbb Z}$, indicating that the vortex solutions constructed below are characterized by two winding numbers, say $(N,M)$.

To search for vortex solutions, it is possible to work with the usual rotationally symmetric ansatz \cite{Manton}, which in our case may be written out as  
\begin{align}
&a^{(1)}_r=a^{(2)}_r=0\,,\quad a^{1}_\theta := a^{(1)}_{\theta}(r) \,,\quad a^2_\theta := a^{(2)}_\theta(r) \,, \nn \\
&\varphi=\zeta(r) e^{iN\theta}\,,\,\,\, \chi=\eta(r) e^{iM\theta} \,, \quad \varphi_3 = \rho(r) \,,\quad \chi_3 = \sigma(r)\,, \quad \psi = \tau(r) \,,
\label{comfields}
\end{align} 
where the cartesian coordinates $(y_1, y_2)$ are replaced by the polar variables $(r \,, \theta)$. With this ansatz the action reads 
\begin{align}
S = 2 \pi \int dr\bigg(\frac{1}{9g^2r}({a_\theta^1}^\prime{a_\theta^1}^\prime+{a_\theta^2}^\prime{a_\theta^2}^\prime+{a_\theta^1}^\prime{a_\theta^2}^\prime)+\frac{2r}{3}({\zeta^\prime}^2+{\eta^\prime}^2)+\frac{2}{3r}(N+a^1_\theta)^2\zeta^2\nonumber
 \\+\frac{2}{3r}(M+a_\theta^2)^2\eta^2+\frac{r}{4}({\rho^\prime}^2+{\sigma^\prime}^2+{\tau^\prime}^2)-\frac{r}{6}(\rho^\prime{\sigma}^\prime+{\rho}^\prime{\tau}^\prime+{\sigma}^\prime{\tau}^\prime) \nonumber
\\+\frac{4r}{3\tilde{g}^2}\bigg((1-\zeta^2-\eta^2)+\frac{3}{8}(\rho^2+\sigma^2+\tau^2)-\rho+\sigma-\frac{1}{4}(\rho\sigma+\rho\tau+\sigma\tau)\nonumber
\\+\zeta^4+\eta^4-\zeta^2\eta^2+\zeta^2(\rho^2+\rho)+\eta^2(\sigma^2-\sigma)\bigg)\bigg)\,,
\label{action3}
\end{align} 
where primes are denoting the derivatives with respect to $r$. 

Euler-Lagrange equations for the fields are 
\begin{align}
 &\zeta^{\prime\prime}+\frac{\zeta^\prime}{r}-\bigg(\frac{1}{r^2}(N+a_\theta^1)^2+\frac{2}{\tilde {g}^2}(-1+2\zeta^2-\eta^2+\rho^2+\rho)\bigg)\zeta=0\,,\nonumber \label{Euler-Lagrange1}
 \\&\eta^{\prime\prime}+\frac{\eta^\prime}{r}-\bigg(\frac{1}{r^2}(M+a_\theta^2)^2+\frac{2}{\tilde {g}^2}(-1+2\eta^2-\zeta^2+\sigma^2-\sigma)\bigg)\eta=0\,,\nonumber
\\&{a^1_\theta}^{\prime\prime}-\frac{{a^1_\theta}^\prime}{r}+\frac{1}{2}{a^2_\theta}^{\prime\prime}-\frac{{a^2_\theta}^\prime}{2r}-6g^2(N+a_\theta^1)\zeta^2=0\,,\nonumber
\\&{a^2_\theta}^{\prime\prime}-\frac{{a^2_\theta}^\prime}{r}+\frac{1}{2}{a^1_\theta}^{\prime\prime}-\frac{{a^1_\theta}^\prime}{2r}-6g^2(M+a_\theta^2)\eta^2=0\,,
 \\&\rho^{\prime\prime}+\frac{\rho^\prime}{r}-\frac{\sigma^\prime+\tau^\prime}{3r}-\frac{\sigma^{\prime\prime}+\tau^{\prime\prime}}{3}-\frac{2\rho}{\tilde{g}^2}+\frac{8}{3\tilde{g}^2}+\frac{2}{3\tilde{g}^2}(\sigma+\tau)-\frac{8}{3\tilde{g}^2}\zeta^2(2\rho+1)=0\,,\nonumber
\\&\sigma^{\prime\prime}+\frac{\sigma^\prime}{r}-\frac{\rho^\prime+\tau^\prime}{3r}-\frac{\rho^{\prime\prime}+\tau^{\prime\prime}}{3}-\frac{2\sigma}{\tilde{g}^2}-\frac{8}{3\tilde{g}^2}+\frac{2}{3\tilde{g}^2}(\rho+\tau)-\frac{8}{3\tilde{g}^2}\eta^2(2\sigma-1)=0\,,\nonumber
\\&\tau^{\prime\prime}+\frac{\tau^\prime}{r}-\frac{\rho^\prime+\sigma^\prime}{3r}-\frac{\rho^{\prime\prime}+\sigma^{\prime\prime}}{3}-\frac{2\tau}{\tilde{g}^2}+\frac{2}{3\tilde{g}^2}(\rho+\sigma)=0\,.\nonumber
\end{align} 
We do not know any analytic solutions to these coupled non-linear differential equations. However, we can construct the solutions profiles for small and large $r$. For $r\rightarrow 0$, series solutions give 
\begin{align}
\zeta=\zeta_0r^N+\mathit{O}(r^{N+2})&\,,\quad \eta=\eta_0r^M+\mathit{O}(r^{M+2})\,,\quad a_\theta^1=a_0^{(1)}r^2+\mathit{O}(r^4)\,,\quad a_\theta^2=a_0^{(2)}r^2+\mathit{O}(r^4)\nonumber
\\&\rho=\rho_0+\mathit{O}(r^2)\,,\quad \sigma=\sigma_0+\mathit{O}(r^2)\,,\quad \tau=\tau_0+\mathit{O}(r^2)\,,
\label{smallr}
\end{align}
where $\zeta_0\,,\eta_0\,,a_0^{(1)}\,,a_0^{(2)}\,,\rho_0\,,\sigma_0\,,\tau_0$ are constants. 

For large $r$, we first note that the asymptotic behavior of fields are enforced by the requirement of the finiteness of the action for the vortex type solutions. We have $\zeta(r)\rightarrow1\,,\eta(r)\rightarrow1\,,a_\theta^1(r)\rightarrow N\,,a_\theta^2(r)\rightarrow M\,,\rho(r)\rightarrow 0\,,\sigma(r)\rightarrow 0\,,\tau(r)\rightarrow 0$ as $r\rightarrow \infty$, where the integers $N$ and $M$ are the winding numbers of the vortex configuration.  
In order to obtain the profiles for large $\ell$, we can consider the small fluctuations about these limiting values and write $\zeta=1-\delta\zeta\,,\eta=1-\delta\eta\,,a_\theta^1=-N+\delta a^1 \,,a_\theta^2=-M+\delta a^2$. Assuming that $(\frac{\delta a_\theta^1}{r})^2$ and $(\frac{\delta a_\theta^2}{r})^2$ are subleading compared to $\delta\zeta\,,\delta\eta\,,\rho\,,\sigma\,,\tau$, the Euler-Lagrange equations (\ref{Euler-Lagrange1}) become
\begin{align}
 &\delta\zeta^{\prime\prime}+\frac{\delta\zeta^\prime}{r}-\frac{2}{\tilde{g}^2}(4\delta\zeta-\rho-2\delta\eta)=0\,,\quad \delta\eta^{\prime\prime}+\frac{\delta\eta^\prime}{r}-\frac{2}{\tilde{g}^2}(4\delta\eta+\sigma-2\delta\zeta)=0\nonumber
 \\&{\delta a^1}^{\prime\prime}-\frac{{\delta a^1}^\prime}{r}+4g^2{\delta a^2}-8g^2\delta a^1=0\,,\quad{\delta a^2}^{\prime\prime}-\frac{{\delta a^2}^\prime}{r}+4g^2{\delta a^1}-8g^2\delta a^2=0\,,\nonumber
  \\&{\rho^{\prime\prime}}+\frac{\rho^\prime}{r}-\frac{10}{\tilde{g}^2}\rho-\frac{4}{\tilde{g}^2}\sigma+\frac{8}{\tilde{g}^2}\delta\zeta-\frac{4}{\tilde{g}^2}\delta\eta=0\,,
  \\&{\sigma^{\prime\prime}}+\frac{\sigma^\prime}{r}-\frac{10}{\tilde{g}^2}\sigma-\frac{4}{\tilde{g}^2}\rho-\frac{8}{\tilde{g}^2}\delta\eta+\frac{4}{\tilde{g}^2}\delta\zeta=0\,,\nonumber
  \\&{\tau^{\prime\prime}}+\frac{\tau^\prime}{r}-\frac{2}{\tilde{g}^2}\tau-\frac{4}{\tilde{g}^2}\rho-\frac{4}{\tilde{g}^2}\sigma-\frac{4}{\tilde{g}^2}\delta\eta+\frac{4}{\tilde{g}^2}\delta\zeta=0\,,\nonumber
\end{align}
We can solve these coupled linear differential equations in terms of the modified Bessel functions $K_\alpha$ and find
\begin{align}
\delta\zeta&=A_1K_0(\frac{2\sqrt{2}r}{\tilde g})+A_2K_0(\frac{\sqrt{2}r}{\tilde g})-A_3K_0(\frac{3\sqrt{2}r}{\tilde g})\,,\nn \\ 
\delta\eta&=A_2K_0(\frac{\sqrt{2}r}{\tilde g})+A_3K_0(\frac{3\sqrt{2}r}{\tilde g})+A_4K_0(\frac{2\sqrt{2}r}{\tilde g})\,,\nn \\ 
\rho&=A_2K_0(\frac{\sqrt{2}r}{\tilde g})+3A_3K_0(\frac{3\sqrt{2}r}{\tilde g})-2A_4K_0(\frac{2\sqrt{2}r}{\tilde g})\,, \nn \\
\sigma&=2A_1K_0(\frac{2\sqrt{2}r}{\tilde g})-A_2K_0(\frac{\sqrt{2}r}{\tilde g})+3A_3K_0(\frac{3\sqrt{2}r}{\tilde g})\,, \label{aaa} \\
\tau&=\frac{2}{3}(A_1-A_4)K_0(\frac{2\sqrt{2}r}{\tilde g})+2A_3K_0(\frac{3\sqrt{2}r}{\tilde g})+A_5K_0(\frac{\sqrt{2}r}{\tilde g})\,, \nn \\
{\delta a^1}&=C_1rK_1(2gr)+C_2rK_1(2\sqrt{3}gr)\,, \nn \\
{\delta a^2}&=C_1rK_1(2gr)-C_2rK_1(2\sqrt{3}gr)\,, \nn
\end{align} 
where $A_i\,,i=1\cdots,5$ and $C_j\,,j=1,2$ are constants, which can only be determined numerically. It is easy to see that our assumption that $(\frac{\delta a_\theta^1}{r})^2$ and $(\frac{\delta a_\theta^2}{r})^2$ are subleading to $\delta\zeta\,,\delta\eta\,,\rho\,,\sigma\,,\tau$ can be fulfilled if we take $4g>\sqrt{2}/\tilde g$. A well-known fact is that the field strength and scalars are, respectively, responsible for the repulsive and attractive character of forces between vortices \cite{Manton}.  We find from (\ref{aaa}) that, the field strengths $B^1 := f_{12}^{1} = \frac{1}{r} f_{r \theta}^{1} = \frac{1}{r} \partial_r a_\theta^1$ and $B^2 := f_{12}^{2} = \frac{1}{r} f_{r \theta}^{2} = \frac{1}{r} \partial_r a_\theta^2$ are proportional to $\propto \frac{1}{\sqrt{r}}e^{-2 g r}$ while the scalar fields $\delta\zeta\,,\delta\eta\,,\rho\,,\sigma$ and $\tau$ decay like $\frac{1}{\sqrt{r}}e^{-\frac{\sqrt{2}}{\tilde{g}} r}$ asymptotically. Thus these vortices attract for $g {\tilde g} > \frac{\sqrt{2}}{2}$ and particularly for the case $g {\tilde g} =1$ needed for the standard Yang-Mills (\ref{stym}), and they repel in the parameter interval $\frac{\sqrt{2}}{4} < g {\tilde g} < \frac{\sqrt{2}}{2}$. From the asymptotic profiles of the fields, we can not immediately conclude the presence of BPS solutions at the point $g {\tilde g} = \frac{\sqrt{2}}{2}$ of the parameter space, where there appears to be a change between attractive and repulsive nature of forces between vortices. In fact, we do not find any BPS equations from (\ref{action2}) at this point of the parameter space, while as we shall see in the next subsection, $g {\tilde g} = 1$ is a critical point at which BPS vortices are found as $\ell \rightarrow \infty$ and $a \rightarrow \infty$.

\subsection{Case $ii)$}

Taking the limit $a\rightarrow \infty$ is equivalent to enforcing the constraint $\Phi_a\Phi_a+\ell(\ell+1)=0$. It can be easily seen from (\ref{constraint}) that this constraint can only be fulfilled by setting $R_1=0\,,R_2=0$ and $R_3=0$. Using these three conditions, we can solve $\varphi_3\,,\chi_3$ and $\psi$ in terms of $|\varphi|$ and $|\chi|$ in powers of $\frac{1}{\ell}$. Substituting back into the action should then give us an action with only two complex scalars $\varphi$ and $\chi$. To leading non-vanishing order in powers of $\frac{1}{\ell}$ , we find that 
\begin{align}
 \psi&=\frac{1}{2\ell}(1-|\varphi|^2)+\frac{1}{2\ell}(1-|\chi|^2)+\mathit{O(\frac{1}{\ell^2})}\,,\nonumber
 \\\varphi_3&=-\frac{3}{4\ell^2}(1-|\varphi|^2)-\frac{2\ell+1}{4\ell^2}(1-|\chi|^2)+\mathit{O(\frac{1}{\ell^3})}\,,\label{realscalars}
 \\\chi_3&=\frac{1}{4\ell^2}(1-|\chi|^2)-\frac{2\ell+1}{4\ell^2}(1-|\varphi|^2)+\mathit{O(\frac{1}{\ell^3})}\,.\nonumber
\end{align}
Substituting from (\ref{realscalars}) for $\varphi_3\,,\chi_3\,,\psi$, expanding $\ell$ dependent coefficients to order $\frac{1}{\ell^2}$, the action (\ref{notac}) takes the form
\begin{align}
 S&=\int  d^2 y \, \, \bigg(\frac{1}{18g^2}(1+\frac{1}{2\ell}-\frac{3}{4\ell^2})f^{(1)}_{\mu\nu}{f^{(1)}}^{\mu\nu}+\frac{1}{18g^2}(1-\frac{1}{2\ell}-\frac{1}{4\ell^2})f^{(2)}_{\mu\nu}{f^{(2)}}^{\mu\nu} \nn
 \\&+\frac{1}{18g^2}(1-\frac{1}{\ell^2})f^{(1)}_{\mu\nu}{f^{(2)}}^{\mu\nu}+\frac{2}{3}(1-\frac{1}{2\ell^2})(|D_\mu\varphi|^2+|D_\mu\chi|^2)\nn
 \\&+\frac{1}{6\ell^2}\big((\partial_\mu|\varphi|^2)^2+(\partial_\mu|\chi|^2)^2+\partial_\mu|\varphi|^2\partial_\mu|\chi|^2\big)+\frac{1}{\tilde{g}^2}\bigg(\frac{4}{3}(1+\frac{1}{4\ell^2})-\frac{4}{3}(1-\frac{1}{\ell}+\frac{1}{\ell^2})|\varphi|^2\nn
 \\&-\frac{4}{3}(1+\frac{1}{\ell}-\frac{1}{\ell^2})|\chi|^2-\frac{4}{3}(1+\frac{3}{4\ell^2})|\varphi|^2|\chi|^2+\frac{4}{3}(1-\frac{1}{2\ell}+\frac{1}{2\ell^2})|\varphi|^4+\frac{4}{3}(1+\frac{1}{2\ell}-\frac{1}{2\ell^2})|\chi|^4\nn
 \\&+\frac{1}{3\ell^2}(|\varphi|^4|\chi|^2+|\chi|^4|\varphi|^2)\bigg)\bigg)\,,
 \label{action5}
\end{align}
where we wrote 
\begin{align}
h_{\mu\nu}=-\frac{2}{3}(\frac{1}{\ell}-\frac{1}{2\ell^2})(f^{(1)}_{\mu\nu}+f^{(2)}_{\mu\nu})\,,
\end{align} 
which follows from the equation of motion of $b_\mu$ at the $\frac{1}{\ell^2}$ order. 

For this case too, we make the rotationally symmetric vortex solution ansatz (\ref{comfields}) and find the action to take the form
\begin{align}
S &=2\pi\int dr\bigg(\frac{1}{9g^2r}(1+\frac{1}{2\ell}-\frac{3}{4\ell^2}){a_{\theta}^1}^\prime{a_{\theta}^1}^\prime+\frac{1}{9g^2r}(1-\frac{1}{2\ell}-\frac{1}{4\ell^2}){a_{\theta}^2}^\prime{a_{\theta}^2}^\prime+\frac{1}{9g^2r}(1-\frac{1}{\ell^2}){a_{\theta}^1}^\prime{a_{\theta}^2}^\prime\nonumber
 \\&+\frac{2}{3}(1-\frac{1}{2\ell^2})(r{\zeta^\prime}^2+\frac{(N+a_\theta^1)^2}{r}\zeta^2+r{\eta^\prime}^2+\frac{(M+a_\theta^2)^2}{r}\eta^2)+\frac{r}{6\ell^2}\big(4{\zeta^\prime}^2\zeta^2+4{\eta^\prime}^2\eta^2+4\zeta^\prime\zeta\eta^\prime\eta\big)\nonumber
 \\&+\frac{1}{\tilde{g}^2}\bigg(\frac{4r}{3}(1+\frac{1}{4\ell^2})-\frac{4r}{3}(1-\frac{1}{\ell}+\frac{1}{\ell^2})\zeta^2-\frac{4r}{3}(1+\frac{1}{\ell}-\frac{1}{\ell^2})\eta^2-\frac{4r}{3}(1+\frac{3}{4\ell^2})\zeta^2\eta^2\nonumber
 \\&+\frac{4r}{3}(1-\frac{1}{2\ell}+\frac{1}{2\ell^2})\zeta^4+\frac{4r}{3}(1+\frac{1}{2\ell}-\frac{1}{2\ell^2})\eta^4+\frac{r}{3\ell^2}(\zeta^4\eta^2+\eta^4\zeta^2)\bigg)\bigg) \,.
 \label{actionii}
\end{align}
Equation of motions for the fields $\zeta\,,\eta\,, a_\theta^1\,,a_\theta^2$ after a straightforward calculation are given in the appendix. Profiles of these fields around $r=0$ are the same as in the previous case (\ref{smallr}). 

For large $r$, it is easy to find the linearized equations for the fluctuations about the vacuum values. We write as before  $\zeta=1-\delta\zeta\,,\eta=1-\delta\eta\,,a_\theta^1=-N+\delta a^1\,,a_\theta^2=-M+\delta a^2$ and we obtain the equations
\begin{align}
 &\delta\zeta^{\prime\prime}+\frac{\delta\zeta^\prime}{r}-\frac{2}{\tilde{g}^2}(4-\frac{2}{\ell}+\frac{2}{\ell^2})\zeta+\frac{2}{\tilde{g}^2}(2+\frac{1}{2\ell^2})\eta=0\,,\nonumber
 \\&\delta\eta^{\prime\prime}+\frac{\delta\eta^\prime}{r}-\frac{2}{\tilde{g}^2}(4+\frac{2}{\ell}-\frac{2}{\ell^2})\eta+\frac{2}{\tilde{g}^2}(2+\frac{1}{2\ell^2})\zeta=0\,,\nonumber
 \\&{\delta a^1}^{\prime\prime}-\frac{{\delta a^1}^\prime}{r}-2g^2(4-\frac{2}{\ell}+\frac{1}{\ell^2})\delta a^1+2g^2(2-\frac{1}{\ell^2})\delta a^2=0\,,\nonumber
 \\&{\delta a^2}^{\prime\prime}-\frac{{\delta a^2}^\prime}{r}-2g^2(4+\frac{2}{\ell}-\frac{1}{\ell^2})\delta a^2+2g^2(2-\frac{1}{\ell^2})\delta a^1=0\,.
\end{align}
Solutions for these equations are given in terms of modified Bessel functions $K_n$: 
\begin{align}
 \delta\zeta&=E_1(-1+\frac{1}{\ell}+\frac{3}{2\ell^2})K_0\big(\frac{\sqrt{12+3/\ell^2}r}{\tilde{g}}\big)+E_2(1+\frac{1}{\ell}-\frac{1}{2\ell^2})K_0\big(\frac{\sqrt{4-3/\ell^2}r}{\tilde g}\big)\,,\nonumber
 \\\delta\eta&=E_1K_0\big(\frac{\sqrt{12+3/\ell^2}r}{\tilde{g}}\big)+E_2K_0\big(\frac{\sqrt{4-3/\ell^2}r}{\tilde g}\big)\,,\nonumber
 \\\delta a^1&=F_1(-1+\frac{1}{\ell}-\frac{1}{\ell^2})rK_1(2\sqrt{3}gr)+F_2(1+\frac{1}{\ell})rK_1(2gr)\,,\nonumber
 \\\delta a^2&=F_1 r K_1(2\sqrt{3}gr)+F_2 r K_1(2gr) \,,
\end{align} where $E_1\,,E_2\,,F_1\,,F_2$ are constants. Here, we can also define the parameter intervals for the attractive and repulsive behaviour of forces between the vortices. It is easy to see that for $g\tilde{g}>\frac{\sqrt{4-3/\ell^2}}{2}$, the field strengths decay faster than the scalar fields, so we have attractive vortices. On the other hand, for $\frac{\sqrt{4-3/\ell^2}}{4} < g {\tilde g} < \frac{\sqrt{4-3/\ell^2}}{2}$ we have repulsive forces between the vortices.

As $\ell\rightarrow \infty$ the action (\ref{action5}) at the critical point $g {\tilde g} =1$ becomes  
\begin{align}
S=\int d^2y \, \, \frac{1}{18 g^2} \big (f^{(1)}_{\mu\nu}{f^{(1)}}^{\mu\nu}+f^{(2)}_{\mu\nu}{f^{(2)}}^{\mu\nu}+f^{(1)}_{\mu\nu}{f^{(2)}}^{\mu\nu} \big)+\frac{2}{3}(|D_\mu\varphi|^2+|D_\mu\chi|^2)\nonumber \\
+\frac{2}{3} g^2 \bigg ((|\varphi|^2+\varphi_3 -1)^2 + (|\chi|^2 - \chi_3 -1)^2 + (|\varphi|^2-|\chi|^2)^2 \bigg) \,,
\end{align} 
In this case we may express the action in the form
\begin{align}
S =\int d^2y \, \, \frac{1}{18g^2}\big(B^1+ 2g^2 (2|\varphi|^2-|\chi|^2-1)\big)^2+\frac{1}{18 g^2} \big(B^2+ 2g^2(2|\chi|^2-|\varphi|^2-1)\big)^2\nonumber
 \\+\frac{1}{18g^2}\big(B^1+B^2+ 2g^2(|\varphi|^2+|\chi|^2-2)\big)^2 +\frac{2}{3}\big(\overline{D_1\varphi}-i\overline{D_2\varphi}\big)\big(D_1\varphi + i D_2\varphi\big)\nonumber
 \\+\frac{2}{3}\big(\overline{D_1\chi}-i\overline{D_2\chi}\big)\big(D_1\chi+iD_2\chi\big)+\frac{2}{3}(B^1+B^2)\nonumber
 \\-\frac{2i}{3}\big(\partial_1(\overline{\varphi}D_2\varphi)-\partial_2(\overline{\varphi}D_1\varphi)\big)-\frac{2i}{3}\big(\partial_1(\overline{\chi}D_2\chi)-\partial_2(\overline{\chi}D_1\chi)\big)\,,\label{energy}
\end{align}
where $B^1=f_{12}^1\,,B^2=f_{12}^2$ as we have noted previously. The last two terms in (\ref{energy}) vanish as they can be expressed as line integrals around a circle at infinity. Noting that the fluxes of $B^1$ and $B^2$ are $2\pi N$ and $2\pi M$ respectively, $N, M$ being the winding numbers of the vortex configuration, we see that the action is bounded from below with $S \geq \frac{4}{3}\pi(N+M)$. This bound is saturated, when the fields satisfy the BPS equations:
\begin{align}
& D_1\varphi+iD_2\varphi=0\,,\quad B^1+ 2g^2 (2|\varphi|^2 -|\chi|^2 -1) = 0\,,\nonumber
\\&D_1\chi+iD_2\chi=0\,,\quad B^2 + 2g^2 (2|\chi|^2-|\varphi|^2-1) = 0 \,.
\end{align}
These equations give a particular generalization of the BPS equations for the abelian Higgs model \cite{Manton}. In fact, these equation appear to be formally the same as the self dual instanton equations for the $SU(3)$ Yang-Mills theory with cylindrical symmetry studied by Bais and Weldon \cite{Bais}. There is a clear distinction between the two however; the latter are in the context of Yang-Mills theories over ${\mathbb R}^4$ and the cylindrically symmetric ansatz essentially dimensionally reduces that theory to an abelian Higgs type model over ${\mathbb H}^2$, with the $SU(3)$ instanton solutions being characterized by a Pontryagin index, which is given as the sum of the two winding numbers of the abelian Higgs type model over ${\mathbb H^2}$ with $U(1) \times U(1)$ gauge symmetry, while our BPS equations are obtained for $U(1) \times U(1)$ abelian Higgs type model over ${\mathbb R^2}$. 

\section{Generalization of $SU(2)$-equivariant gauge fields for $U(n)$ gauge theory}\label{Generalization}

Now, we briefly indicate how the results of section 3 generalizes to $U(n)$ gauge theories over ${\cal M}\times S_F^2$. For this purpose we write the symmetry generators $\omega_a$
\begin{align}
 \omega_a=X_a^{(2\ell+1)}\otimes \bm{1}_n-\bm{1}^{(2\ell+1)}\otimes i\tilde{\Sigma}^k_a\,,
\end{align} where $\tilde{\Sigma}^k_a$ are spin $k$ irreducible representation of $SU(2)$ with $n = 2k+1$. Thus, the $SU(2)$ IRR content of $\omega_a$ is
\begin{align}
\ell\otimes k=(\ell+k)\oplus (\ell+k-1)\oplus \cdots \oplus |\ell-k|\,,
\end{align} and the IRR content of the adjoint action of $\omega_a$ can be found to be
\begin{align}
 [\ell\otimes k]^{\otimes 2}={\bm{(2k+1)}} 0 \oplus {\bm {(6k+1)}} 1 \oplus \cdots\,.
 \label{pole}
\end{align} 
This decomposition means that under the adjoint action of $\omega_a$, there are $(2k+1)$ scalars and $(6k+1)$ vectors. It indicates that with our symmetry constraints (\ref{symcons}), the set of solutions to $A_\mu$ should be $(2k+1)$-dimensional while the set of the solutions to $A_a$ should be $(6k+1)$-dimensional. It is possible to find the parametrization of $A_\mu$ by using the following rotational invariants 
\begin{align}
 \bm{1}_{(2\ell+1)(2k+1)}\,,\quad \tilde{\Sigma}^k_aX_a\,,\quad (\tilde{\Sigma}^k_aX_a)^2\,,\quad (\tilde{\Sigma}^k_aX_a)^3\,,\quad \cdots,\quad(\tilde{\Sigma}^k_aX_a)^{2k}\,.\label{poleinv}
\end{align}
We may recall that the adjoint representation of $SU(n)$ is $n^2-1$ dimensional and decomposes under the $SU(2)$ IRRs as
\begin{align}
n^2-1=\oplus \sum_{j=1}^{n-1}(2j+1)\,.
\end{align} 
This is a multipole expansion starting with the dipole term and going up to the $(n-1)^{\text{th}}$-pole term. Tjus, considering that we may construct one rotational invariant per multipole term, then together with the identity we have $n=2k+1$ rotational invariants as we have already inferred from (\ref{pole}). The invariants listed in (\ref{poleinv}) may be expressed in terms of the appropriate multipole tensors and can further be combined into idempotents as we given in (\ref{idem}) for the case of $k=1$ and the vectors can be obtained subsequently. 

\section{Equivariant field modes over other vacuum configurations}

It is possible to investigate the structure of equivariant fields over other fuzzy vacuum configurations. One such case of particular interest is the vacuum configuration 
\begin{align}
S_F^{2 \, Int} := S_F^2 (\ell) \oplus S_F^2 (\ell) \oplus  S_F^2 \left ( \ell + \frac{1}{2} \right ) \oplus S_F^2 \left ( \ell - \frac{1}{2} \right ) \,,
\label{concen}
\end{align} studied in \cite{Seckinson}.\footnote{Note that, in this case $V_2(\Phi)$ term is omitted from the action (\ref{action}). Nevertheless, it is possible to impose it as a constraint as discussed in \cite{Seckinson}.}. There, the structure of this vacuum was revealed by performing the field redefinition
\begin{align}
\Phi_a = \phi_a + \Gamma_a \,, \quad \Gamma_a = -\frac{i}{2}\Psi^\dagger {\tilde \tau}_a \Psi  \,,
\label{tau}
\end{align}
where
\begin{align}
\Psi = \left (
\begin{array}{c}
\Psi_1 \\
\Psi_2
\end{array}
\right ) \,,\quad \Psi_\alpha \in \text{Mat}({\cal N})\,,\quad \alpha=1,2\,,
\label{doublets}
\end{align}
is a doublet of the global $SU(2)$ symmetry of the action (\ref{action}). In (\ref{tau}) and (\ref{doublets}), $\phi_a$, $\Psi_\alpha$ and $\Gamma_a$ are all transforming adjointly under $SU(\cal{N})$ and ${\tilde \tau}_a=\tau_a \otimes 1_{{\cal N}}$ with $\tau_a$ being the Pauli matrices. We note that $\phi_a \,,(a=1,2,3)$ have $3N^2$ real degrees of freedom while $\Psi$ has $4N^2$ real degrees of freedom in total. However, what enters into the definition of $\Gamma_a$ are the equivalence classes $\Psi \sim U\Psi$, $U\in SU({\cal N})$, as it can readily be observed that $\Gamma_a$ are invariant under the left action $U \Psi$ of $SU({\cal N})$ on $\Psi$. It is thus clear that $\Gamma_a\,(a=1,2,3)$ have in total $4N^2-N^2=3N^2$ degrees of freedom as $\phi_a$'s do and (\ref{tau}) is indeed a reparametrization of the fields $\Phi_a$ \cite{Gonul-Seckin}.

Using (\ref{tau}), we see that up to gauge transformations (\ref{gt}) the vacuum configuration is given as
\begin{align}
\Phi_a = ( X_a^{(2 \ell + 1)} \otimes {\bf 1}_4 \otimes {\bf 1}_n ) + ( {\bf 1}_{2 \ell + 1} \otimes \Gamma_a^0 \otimes {\bf 1}_n )  \,,\label{IRRvacua}
\end{align} 
where $ \Gamma_a^0 = - \frac{i}{2} \psi^\dagger \tau_a \psi$ are $4 \times 4$ matrices and the two-component spinor $\Psi^0 \equiv \psi$ is taken as
\begin{align}
\psi = \left (
\begin{array}{c}
\psi_1 \\
\psi_2
\end{array}
\right )
:=
\left (
\begin{array}{c}
b_1 \\
b_2
\end{array}
\right ) \,,
\label{solspi}
\end{align} 
and where $b_\alpha \,, b_\alpha^\dagger$ are two sets of fermionic annihilation-creation operators which span the $4$-dimensional Hilbert space with the basis vectors
\begin{align}
|n_1 \,, n_2 \rangle \equiv (b_1^\dagger)^{n_1} (b_2^\dagger)^{n_2} | 0 \,, 0 \rangle \,, \quad n_1 \,, n_2 = 0 \,,1 \,.
\label{eq:basis1}
\end{align}
$SU(2)$ IRR content of $\Gamma_a^0$ is
\begin{align}
0_{\bm 0} \oplus 0_{\bm 2} \oplus \frac{1}{2} \,,
\label{eq:irrdec}
\end{align} 
where $0_{\bm 0}\,,0_{\bm 2}$ stand for the two inequivalent singlets. These two singlets are distinguished by the eigenvalues of the number operator $N=b_\alpha^\dagger b_\alpha$ which take the values $0$ and $2$, respectively. It is easy to see that the projections to the singlet and doublet subspaces respectively may be found on these representations as
\begin{align}
P_0 &=1 - N + 2 N_1 N_2 \,,\nonumber
\\P_{0_{\bm 0}} &=- \frac{1}{2} (N - 2 ) P_0 = 1 - N + N_1 N_2 \,, \nonumber \\
P_{0_{\bm 2}} &= \frac{1}{2} N P_0 =  N_1 N_2 =  - \frac{1}{2} N +\frac{1}{2} P_{\frac{1}{2}} \,,\nonumber
\\P_{\frac{1}{2}} &= N - 2 N_1 N_2  \,,
\end{align} where $N=N_1+N_2\,,N_1=b_1^\dagger b_1\,,N_2=b_2^\dagger b_2$. 

$SU(2)$ IRR content of vacuum configuration (\ref{IRRvacua}) can be derived from the Clebsch-Gordan decomposition as
\begin{align}
\ell \otimes \left (0_{\bm 0} \oplus 0_{\bm 2} \oplus \frac{1}{2} \right) \equiv  \ell \oplus \ell \oplus \left ( \ell + \frac{1}{2} \right )\oplus \left ( \ell - \frac{1}{2} \right )\,, \quad \ell \neq 0 \,.
\label{eq:1deco1}
\end{align} 
This indicates that the vacuum configuration (\ref{concen}) can be interpreted as a direct sum of four concentric fuzzy spheres as it has been already discussed in \cite{Seckinson}. In that article low energy structure of $U(2)$ gauge theory over ${\cal M}\times S_F^{2 \, Int}$ was investigated in detail. Here, our aim is to consider the $U(3)$ gauge theory over ${\cal M}\times S_F^{2 \, Int}$ and construct the $SU(2)$ equivariant gauge fields characterizing its low energy behaviour. In order to determine the latter, we choose the $SU(2)$ symmetry generators $\omega_a$ as
\begin{align}
 \omega_a =& (X_a^{(2 \ell + 1)} \otimes {\bf 1}_4 \otimes {\bf 1}_3) + ({\bf 1}_{2 \ell + 1} \otimes \Gamma_a^0 \otimes {\bf 1}_3)
- ({\bf 1}_{2 \ell + 1} \otimes {\bf 1}_4 \otimes i \Sigma_a)  \nonumber \\
\quad =:& X_a + \Gamma_a^0 - i \Sigma_a \nonumber  
\\=:&D_a - i \Sigma_a \,, \quad  \quad \omega_a \in u(2\ell+1) \otimes u(4) \otimes u(3) \,,
\end{align} 
and they satisfy (\ref{omega}). $\omega_a$ carries a direct sum of IRRs of $SU(2)$, which is given as 
\begin{align}
\bigg( \ell \oplus \ell \oplus \left ( \ell + \frac{1}{2} \right )\oplus \left ( \ell - \frac{1}{2} \right )\bigg)\otimes 1&\equiv {\bf{2}}\bigg((\ell-1)\oplus \ell\oplus (\ell+1)\bigg)\oplus {\bf{2}}\bigg( \big( \ell + \frac{1}{2} \big )\oplus \big( \ell - \frac{1}{2} \big )\bigg)\nonumber
\\ \quad \quad & \oplus (\ell -\frac{3}{2})\oplus (\ell +\frac{3}{2})\,.
\label{rhs}
\end{align} 
Projections to the representations appearing in the r.h.s of (\ref{rhs}) is given in the table below,
\begin{table}\centering
\begin{tabular}{c | c }
    Projector & Representation \\ \hline
    \\
    $\Pi_{0_{\bm 0}}= {\bf 1}_{2 \ell + 1} \otimes P_{0_{\bm 0}} \otimes {\bm 1}_3$ & $(\ell-1)\oplus \ell\oplus (\ell+1)$ 
    \\    $\Pi_{0_{\bm 2}}= {\bf 1}_{2 \ell + 1} \otimes  P_{0_{\bm 2}} \otimes {\bm 1}_3 $ & $(\ell-1)\oplus \ell\oplus (\ell+1)$ 
    \\
    $ \Pi_+ = \frac{1}{2} (i Q_I + \Pi_{\frac{1}{2}})$ & $(\ell -\frac{1}{2})\oplus  (\ell +\frac{1}{2})\oplus  (\ell +\frac{3}{2})$ \\
     $\Pi_- = \frac{1}{2} (-i Q_I + \Pi_{\frac{1}{2}})$ & $(\ell -\frac{3}{2})\oplus  (\ell -\frac{1}{2})\oplus  (\ell +\frac{1}{2})$\\
     $\Pi_0 = \Pi_{0_{\bm 0}} + \Pi_{0_{\bm 2}}={\bf 1}_{2 \ell + 1} \otimes P_{0} \otimes {\bm 1}_3$ & ${\bf{2}}\bigg((\ell-1)\oplus \ell\oplus (\ell+1)\bigg)$ \\
     $\Pi_\frac{1}{2}= \Pi_+ + \Pi_- = {\bf 1}_{2 \ell + 1} \otimes P_{\frac{1}{2}} \otimes {\bm 1}_3$ & ${\bf{2}}\bigg( \big( \ell + \frac{1}{2} \big )\oplus \big( \ell - \frac{1}{2} \big )\bigg)\oplus (\ell -\frac{3}{2})\oplus (\ell +\frac{3}{2})$
\end{tabular}
\caption{Projections to the representations appearing in the r.h.s of (\ref{rhs}).}
\end{table}
where
\begin{align}
Q_I=\frac{i}{\frac{1}{2}(\ell+\frac{1}{2})}(X_a\Gamma_a-\frac{1}{4}\Pi_\frac{1}{2})\,,\quad Q_I^2=-\Pi_\frac{1}{2}\,.
\end{align}
$SU(2)$-equivariant gauge fields can be obtained by imposing the symmetry constraints in (\ref{symcons}) and the additional constraint
\begin{align}
\lbrack \omega_a \,, \Psi_\alpha \rbrack &=  \frac{i}{2} ({\tilde \tau}_a)_{\alpha \beta} \Psi_\beta\,.
\label{spicons}
\end{align}
The dimensions of solution spaces for $A_\mu\,,A_a$ and $\Psi_\alpha$ can be derived by the Clebsch-Gordan decomposition of the adjoint action of $\omega_a$. The relevant part of this decomposition is 
\begin{align}
 \bigg[{\bf{2}}\bigg((\ell-1)\oplus \ell\oplus (\ell+1)\bigg)\oplus {\bf{2}}\bigg( \big( \ell + \frac{1}{2} \big )\oplus \big( \ell - \frac{1}{2} \big )\bigg)\oplus (\ell -\frac{3}{2})\oplus (\ell +\frac{3}{2})\bigg]^{\otimes 2}
 \nonumber\\
 \equiv {\bm {22}}\, 0 \oplus {\bm {40}} \, \frac{1}{2} \oplus {\bm {54}} \,1 \oplus \cdots \,. 
 \label{spiexp}
 \end{align}
This simply means that there are $22$ rotationally invariants and $A_\mu$ may be parametrized by these invariants. A suitable set may be listed as the following projectors and ``idempotents" (in the subspace they belong to)
\begin{align}
&\Pi_{0_{\bm 0}}\,,\quad \Pi_{0_{\bm 2}}\,\quad \Pi_+\,,\quad \Pi_-\,\quad iS_1\,,\quad iS_2\,,\quad Q^1_{0_{\bm 0}}=\Pi_{0_{\bm 0}}Q_1\,,\quad Q^2_{0_{\bm 0}}=\Pi_{0_{\bm 0}}Q_2\,,\nonumber
\\& Q^1_{0_{\bm 2}}=\Pi_{0_{\bm 2}}Q_1\,,\quad Q^2_{0_{\bm 2}}=\Pi_{0_{\bm 2}}Q_2\,,\quad Q^1_-\,,\quad Q^2_-\,,\quad Q^1_+\,,\quad Q^2_+\,,\quad Q^1_{+-}\,,\quad Q^2_{-+}\,,\nonumber
\\&{Q_{S}}_{11}=S_1Q_1\,,\quad {Q_{S}}_{12}=S_1Q_2\,,\quad {Q_{S}}_{21}=S_2Q_1\,,\quad {Q_{S}}_{22}=S_2Q_2\,,\quad Q_F\,,\quad Q_H \,,
\label{spinvariant}
\end{align}
where 
\begin{align}
 &Q^1_-=\frac{1}{\ell(2\ell+3)}\bigg((2\ell+1)(\ell+1)\Pi_-Q_1\Pi_--i\Pi_-\bigg)\,,\nonumber
 \\&Q^2_-=\frac{\ell(2\ell+1)}{(\ell+1)(2\ell-1)}\Pi_-Q_2\Pi_-+\frac{(2\ell+1)}{\ell(2\ell-1)(2\ell+3)}\Pi_-Q_1\Pi_--\frac{i}{\ell(\ell+1)(2\ell-1)(2\ell+3)}\Pi_-\,,\nonumber
 \\&Q^1_+=\frac{(2\ell+1)(\ell+1)}{\ell(2\ell+3)}\Pi_+Q_1\Pi_++\frac{(2\ell+1)^2}{(2\ell-1)(2\ell+3)}\Pi_+Q_2\Pi_+-i\frac{(4\ell^3+4\ell^2-\ell+1)}{\ell(2\ell-1)(2\ell+3)}\Pi_+\,,\nonumber
 \\&Q^2_+=\frac{1}{(\ell+1)(2\ell-1)}\bigg(\ell(2\ell+1)\Pi_+Q_2\Pi_+-i\Pi_+\bigg)\,,\nonumber
 \\&Q^1_{+-}=\Pi_+Q_1\Pi_--i\Pi_\frac{1}{2}+2i\Pi_+\,,\quad Q^2_{-+}=\Pi_-Q_2\Pi_+-i\Pi_\frac{1}{2}+2i\Pi_-\,,\nonumber
 \\&S_i = {\bm 1}_{2 \ell+1} \otimes s_i \otimes {\bm 1}_2 \,, \quad 
s_i =
\left ( 
\begin{array}{cc}
\sigma_i & 0_2 \\
0_2 & 0_2
\end{array}
\right ) \,, \quad i = 1 \,, 2 \,,
\end{align}
and 
\begin{align}
&Q_F=\frac{1}{3}\Gamma_a\Sigma_a-2i(\Gamma_a\Sigma_a)^2-i\frac{4}{3}\Pi_\frac{1}{2}\,,\nonumber
\\&Q_H=\frac{4(2\ell+1)}{6\ell^2+11\ell+1}Q^\prime-\frac{4(2\ell^2+3\ell)}{6\ell^2+11\ell+1}Q^{\prime\prime}-i\frac{(2\ell-1)(\ell+1)}{6\ell^2+11\ell+1}\Pi_+-i\frac{3(2\ell-1)(\ell+1)}{6\ell^2+11\ell+1}\Pi_-\nonumber
\\&+i\frac{4\sqrt{4\ell^2+10\ell+2}}{6\ell^2+11\ell+1}\epsilon_{abc}X_a\Gamma_b\Sigma_c+i\frac{16}{6\ell^2+11\ell+1}(\epsilon_{abc}X_a\Gamma_b\Sigma_c)^2\,,\nonumber
\\&Q^\prime=\frac{\ell(2\ell+1)}{(\ell+1)(2\ell-1)}\Pi_-Q_2\Pi_-+\frac{(2\ell+1)^2}{(2\ell-1)(2\ell+3)}\Pi_-Q_1\Pi_--i\frac{4\ell^3+8\ell^2+3\ell-2}{(\ell+1)(2\ell-1)(2\ell+3)}\Pi_-\,,\nonumber
\\&Q^{\prime\prime}=\frac{(2\ell+1)}{(\ell+1)(2\ell-1)(2\ell+3)}\Pi_+Q_2\Pi_++\frac{(2\ell+1)(\ell+1)}{\ell(2\ell+3)}\Pi_+Q_1\Pi_+-i\frac{1}{\ell(\ell+1)(2\ell-1)(2\ell+3)}\Pi_+\,.
\end{align} 
Using Mathematica it is easy to verify that 
\begin{align}
 &(iS_i)^2=-\Pi_0\,,\quad (Q^i_{0_{\bm 0}})^2=-\Pi^i_{0_{\bm 0}}\,,\quad (Q^i_{0_{\bm 2}})^2=-\Pi^i_{0_{\bm 2}}\,,\quad (Q^i_{\pm})^2=-\Pi_{\pm}\,,\quad (Q^1_{+-})^2=-\Pi_\frac{1}{2}\,,\nonumber
 \\&(Q^2_{-+})^2=-\Pi_\frac{1}{2}\,,\quad ({Q_{S}}_{ij})^2=-\Pi_0\,,\quad Q_F^2=-\Pi_\frac{1}{2}\,,\quad Q_H^2=-\Pi_\frac{1}{2}\,, \quad Q^{\prime 2} = - \Pi_- \,, \quad Q^{\prime 2} = - \Pi_+ \,.
\end{align} 

In the equation (\ref{spiexp}), it is seen that under the adjoint action of $\omega_a$, there are $54$ objects which transform as vectors. Using the rotational invariant in (\ref{spinvariant}), we can construct these as follows
\begin{align}
 &\lbrack D_a \,, Q^i_{0_{\bm 0}}\rbrack \,, \quad Q^i_{0_{\bm 0}}\lbrack D_a \,, Q^i_{0_{\bm 0}}\rbrack \,, \quad \lbrace D_a \,, Q^i_{0_{\bm 0}}\rbrace \,, \nonumber\\[0.4em]
&\lbrack D_a \,, Q^i_{0_{\bm 2}}\rbrack \,, \quad Q^i_{0_{\bm 2}}\lbrack D_a \,, Q^i_{0_{\bm 2}}\rbrack \,, \quad \lbrace D_a \,, Q^i_{0_{\bm 2}}\rbrace \,,\nonumber \\[0.4em]
&\lbrack D_a \,, Q^i_- \rbrack \,, \quad Q^i_- \lbrack D_a \,, Q^i_- \rbrack \,, \quad \lbrace D_a \,, Q^i_- \rbrace \,,\nonumber \\[0.4em]
&\lbrack D_a \,, Q^i_+ \rbrack \,, \quad Q^i_+ \lbrack D_a \,, Q^i_+ \rbrack \,, \quad \lbrace D_a \,, Q^i_+\rbrace \,,\nonumber \\[0.4em]
&\lbrack D_a \,, Q_H \rbrack \,, \quad Q_H \lbrack D_a \,, Q_H \rbrack \,, \quad \lbrace D_a \,, Q_H\rbrace \,, \nonumber\\[0.4em]
&\lbrack D_a \,, Q_{F} \rbrack \,, \quad Q_{F} \lbrack D_a \,, Q_{F} \rbrack \,, \quad \lbrace D_a \,, Q_{F} \rbrace \,,  \nonumber\\[0.4em]
&\lbrack D_a \,,{Q_{S}}_{11} \rbrack \,, \quad Q^1_{0} \lbrack D_a \,, {Q_{S}}_{11} \rbrack \,, \quad \lbrace D_a \,, {Q_{S}}_{11} \rbrace \,,  \\[0.4em]
&\lbrack D_a \,,{Q_{S}}_{12} \rbrack \,, \quad Q^2_{0} \lbrack D_a \,, {Q_{S}}_{12} \rbrack \,, \quad \lbrace D_a \,, {Q_{S}}_{12} \rbrace \,, \nonumber \\[0.4em]
&\lbrack D_a \,,{Q_{S}}_{21} \rbrack \,, \quad Q^1_{0} \lbrack D_a \,, {Q_{S}}_{21} \rbrack \,, \quad \lbrace D_a \,, {Q_{S}}_{21} \rbrace \,, \nonumber \\[0.4em]
&\lbrack D_a \,,{Q_{S}}_{22} \rbrack \,, \quad Q^2_{0} \lbrack D_a \,, {Q_{S}}_{22} \rbrack \,, \quad \lbrace D_a \,, {Q_{S}}_{22} \rbrace \,, \nonumber \\[0.4em]
&\lbrack D_a \,, Q^1_{+-} \rbrack \,, \quad Q^1_{\frac{1}{2}} \lbrack D_a \,, Q^1_{+-} \rbrack \,, \quad \lbrace D_a \,, Q^1_{+-}\rbrace \,, \nonumber\\[0.4em]
&\lbrack D_a \,, Q^2_{-+} \rbrack \,, \quad Q^2_{\frac{1}{2}} \lbrack D_a \,,Q^2_{-+} \rbrack \,, \quad \lbrace D_a \,, Q^2_{-+}\rbrace \,,\nonumber \\[0.4em]
&\Pi_{0_{\bm 0}}\omega_a \,, \quad  \Pi_{02}\omega_a \,,  \quad \Pi_- \omega_a \,,  \quad  \Pi_+ \omega_a \,, \quad S_1 \omega_a \,, \quad S_2 \omega_a \,.\nonumber
\end{align} 
Here $Q^1_{0}=\Pi_0Q_1\,,Q^2_{0}=\Pi_0Q_2\,,Q^1_{\frac{1}{2}}=\Pi_{\frac{1}{2}}Q_1\,,Q^2_{\frac{1}{2}}=\Pi_{\frac{1}{2}}Q_2\,,$ and no sum over repeated indices is implied. It is possible to parametrize $A_a$ in terms of these $54$-objects. For the $40$ objects which transform as spinors under the adjoint action of $\omega_a$, we can, for instance, take
\begin{align}
 &\Pi_{0_{\bm 0}}\beta_\alpha Q_{-+} \,, \quad Q^1_{0_{\bm 0}}\beta_\alpha \Pi_- \,, \quad Q^2_{0_{\bm 0}}\beta_\alpha \Pi_- \,, \quad \Pi_{0_{\bm 0}}\beta_\alpha Q_{+-}\,, \quad Q^1_{0_{\bm 0}}\beta_\alpha \Pi_+ \,, \quad Q^2_{0_{\bm 0}}\beta_\alpha \Pi_+\,, \nonumber\\[0.4em]
&Q^1_{0_{\bm 0}}\beta_\alpha Q_{+-} \,,\quad Q^2_{0_{\bm 0}}\beta_\alpha Q_{-+} \,, \quad \Pi_- \beta_\alpha Q^1_{0_{\bm 2}}\,, \quad \Pi_- \beta_\alpha Q^2_{0_{\bm 2}}\,, \quad \Pi_+ \beta_\alpha Q^1_{0_{\bm 2}}\,, \quad \Pi_+ \beta_\alpha Q^2_{0_{\bm 2}}\,, \quad \nonumber \\[0.4em]
&Q^1_{+-} \beta_\alpha \Pi_{0_{\bm 2}}\,, \quad Q^2_{-+} \beta_\alpha \Pi_{0_{\bm 2}}\,, \quad  Q^1_{+-} \beta_\alpha Q^1_{0_{\bm 2}}\,,\quad Q^2_{-+} \beta_\alpha Q^2_{0_{\bm 2}}\,,\quad S_1 \beta_\alpha \Pi_+ \,, \quad S_1 \beta_\alpha \Pi_- \,, \nonumber \\[0.4em]
&\Pi_- \beta_\alpha S_2 \,, \quad  \Pi_+ \beta_\alpha S_2 \,, \quad {Q_{S}}_{11} \beta_\alpha \Pi_+ \,, \quad {Q_{S}}_{11} \beta_\alpha \Pi_- \,,\quad {Q_{S}}_{12} \beta_\alpha \Pi_+ \,, \quad {Q_{S}}_{12} \beta_\alpha \Pi_- \,,\nonumber   \\[0.4em] 
& \Pi_- \beta_\alpha {Q_{S}}_{21} \,, \quad \Pi_- \beta_\alpha {Q_{S}}_{22}\,,\quad \Pi_+ \beta_\alpha {Q_{S}}_{12}\,,\quad \Pi_+ \beta_\alpha {Q_{S}}_{22}
\,,\quad {Q_{S}}_{11} \beta_\alpha Q^1_{+-} \,, \quad {Q_{S}}_{12} \beta_\alpha Q^2_{-+} \,,\nonumber \\[0.4em]
& Q^1_{+-} \beta_\alpha {Q_{S}}_{21} \,,\quad  Q^2_{-+} \beta_\alpha {Q_{S}}_{22} \,,\quad \Pi_{0_{\bm 0}}\beta_\alpha Q^1_{+}\,,\quad \Pi_{0_{\bm 0}}\beta_\alpha Q^2_{-}\,,\quad {Q_{S}}_{11} \beta_\alpha Q^1_{+}\,,\quad {Q_{S}}_{12} \beta_\alpha Q^2_{-}\,,\nonumber  \\[0.4em]
&Q^1_{+} \beta_\alpha {Q_{S}}_{21}\,,\quad Q^2_{-} \beta_\alpha {Q_{S}}_{22}\,,\quad Q^1_{+} \beta_\alpha \Pi_{0_{\bm 2}}\,,\quad Q^2_{-} \beta_\alpha \Pi_{0_{\bm 2}}\,,
\end{align}

Thus, we have determined all the equivariant low energy degrees of freedom for the $U(3)$ gauge theory over ${\cal M} \times S_F^{2 \, Int}$. A few remarks are now in order.
Firstly, we wish to emphasize once again that, from a geometrical point of view the vacuum $S_F^{2 \, Int}$ may be interpreted as stacks of concentric D2-branes with magnetic monopole fluxes and due to this fact it is possible to think of the equivariant gauge field modes that we have found as the modes of the gauge fields living on the world-volume of these D-branes. Let us also stress that the equivariant spinors given above, do not constitute independent degrees of freedom in the $U(3)$ effective gauge theory over ${\cal M}\times S_F^{2\,Int}$. Their bilinears, however, may be constructed to yield the equivariant scalars and vectors. In other words, it is possible to use these equivariant spinor modes to express the ``square roots'' of the equivariant gauge field modes. 

It is possible to explore the dimensional reduction of the $U(3)$ gauge theory over $S_F^{2\,Int}$ or over its projections, such as the monopole bundles $S_F^{2\,\pm}=S_F^2(\ell)\oplus S_F^2(\ell\pm \frac{1}{2})$ with winding numbers $\pm 1$. In this latter case, it easy to observe that the reduced model will yield two decoupled abelian Higgs type model, each carrying $U(1)^{{\otimes} 3}$ as found in section 4 and the vortex solutions determined in section $5$ will be valid within each sector. Dimensional reduction over $S_F^{2\,Int}$ is quite tedious calculation-wise and will not be considered here.

\vskip 2em

{\bf \large Acknowledgements}

This work is supported by the Middle East Technical University under Project No. BAP--01-05-2016-002.

\vskip 1em

\appendices

\section{Details of the Dimensional Reduction over $S_F^2$}

\begin{align}
 \Lambda_1:&=-\frac{2\ell^4+6\ell^3+4\ell^2-\ell-2}{4\ell(2\ell^4+4\ell^3+\ell^2-\ell-1)}P_1+\frac{2\ell^4+2\ell^3-2\ell^2-\ell-1}{4(\ell+1)(2\ell^4+4\ell^3+\ell^2-\ell-1)}P_2+\frac{\omega_c}{2\ell^2+2\ell+1}\,, \nn
 \\\Lambda_2:&=-\frac{4\ell^4+8\ell^3+5\ell^2}{4(2\ell+1)(2\ell^4+4\ell^3+\ell^2-\ell-1)}P_2-\frac{8\ell^5+18\ell^4+11\ell^3+3\ell^2}{4(\ell+1)(2\ell+1)(2\ell^4+4\ell^3+\ell^2-\ell-1)}P_2 \nn
 \\&+\frac{\ell\omega_c}{2\ell^2+2\ell+1}\,, \nn
 \\\Lambda_3:&=\frac{(\ell+1)(8\ell^4+14\ell^3+5\ell^2-3\ell-2)}{4\ell(2\ell+1)(2\ell^4+4\ell^3+\ell^2-\ell-1)}P_1+\frac{(\ell+1)(4\ell^3+4\ell^2+\ell+1)}{4(2\ell+1)(2\ell^4+4\ell^3+\ell^2-\ell-1)}P_2 \nn
 \\&-\frac{(\ell+1)\omega_c}{2\ell^2+2\ell+1}\,, \nn
 \\\Lambda_4:&=-\frac{4\ell^4+10\ell^3+4\ell^2-\ell-2}{4\ell(2\ell+1)^2(2\ell^4+4\ell^3+\ell^2-\ell-1)}P_1+\frac{4\ell^4+6\ell^3-2\ell^2-5\ell-3}{4(2\ell+1)^2(\ell+1)(2\ell^4+4\ell^3+\ell^2-\ell-1)}P_2 \nn
 \\&+\frac{\omega_c}{(2\ell+1)^2}\,, \nn
\\\Lambda_5:&=\frac{2\ell^5+10\ell^4+14\ell^3+3\ell^2-3\ell-2}{2\ell(\ell+1)(2\ell+1)(2\ell^4+4\ell^3+\ell^2-\ell-1)}P_1-\frac{2\ell^4+2\ell^3-\ell^2-\ell-2}{2(2\ell+1)(2\ell^4+4\ell^3+\ell^2-\ell-1)}P_2 \nn
 \\&-\frac{2\omega_c}{2\ell^2+2\ell+1}\,, \nn
\\ \Lambda_6:&=-\frac{2\ell^4+6\ell^3+5\ell^2+\ell-2}{2(2\ell+1)(2\ell^4+4\ell^3+\ell^2-\ell-1)}P_1-\frac{2\ell^5-6\ell^3-\ell^2+3\ell+2}{2\ell(\ell+1)(2\ell+1)(2\ell^4+4\ell^3+\ell^2-\ell-1)}P_2 \nn
 \\&+\frac{2\omega_c}{2\ell^2+2\ell+1}\,, \nn
 \end{align}
 \begin{align}
 \Lambda_7:&=\frac{2\ell^3+6\ell^2+3\ell-3}{2(\ell+1)(2\ell+1)^2(\ell^2+\ell-1)}P_1+\frac{2\ell^3-3\ell+2}{2\ell(2\ell+1)^2(\ell^2+\ell-1)}P_2-\frac{2\omega_c}{(2\ell+1)^2}\,, \nn
 \\\Lambda_8:&=\Lambda_9:=-\Lambda_{13}:=-\frac{1}{(2\ell+1)^2}\,,\quad \Lambda_{10}:=\frac{2\ell^2+3\ell-1}{2(\ell+1)(2\ell+1)}\varphi_3-\frac{1}{2(2\ell+1)}\chi_3+\frac{1}{2\ell+1}\psi\,, \nn
 \\\Lambda_{11}:&=-\frac{2\ell^2+\ell-2}{2\ell(2\ell+1)}\chi_3-\frac{1}{2(2\ell+1)}\varphi_3+\frac{1}{2\ell+1}\psi\,, \nn
 \\\Lambda_{12}:&=\frac{1}{2\ell+1}(-Q_1[X_c,Q_1]-Q_2[X_c,Q_2]-\omega_c+2X_c)
\end{align}
where $P_1:=-Q_1[X_c,Q_1]-i\lbrace X_c,Q_2\rbrace$ and $P_2:=-Q_2[X_c,Q_2]-i\lbrace X_c,Q_1\rbrace$.

\begin{align}
 \alpha_1&=\frac{4(\ell^2+\ell-1)^2(\ell^2+\ell+1)}{3\ell^3(\ell+1)^3}\,,\quad \alpha_2=\frac{4(2\ell^4+5\ell^3+\ell^2-\ell+3)}{3(\ell+1)^3(2\ell+1)}\,,\nonumber
 \\\alpha_3&=\frac{4(2\ell^4+3\ell^3-2\ell^2-4\ell+2)}{3\ell^3(2\ell+1)}\,,\quad \alpha_4=\alpha_5\nonumber \\\alpha_5&=\frac{2(-3\ell^8-12\ell^7-14\ell^6+13\ell^4+12\ell^3+16\ell^2+12\ell-12)}{3\ell^3(\ell+1)^3(2\ell+1)^2}\,,\nonumber
 \\\alpha_6&=\frac{4(4\ell^7+10\ell^6+2\ell^5-2\ell^4-3\ell^3-15\ell^2+4)}{3\ell^3(\ell+1)^2(2\ell+1)^2}\,,\nonumber
 \\\alpha_7&=\frac{4(4\ell^7+18\ell^6+26\ell^5+2\ell^4-35\ell^3-28\ell^2+7\ell+6)}{3\ell^2(\ell+1)^3(2\ell+1)^2}\,,\nonumber
 \\\alpha_8&=\frac{4(\ell^6+3\ell^5+15\ell^4+25\ell^3-30\ell^2-42\ell+24)}{3\ell^2(\ell+1)^2(2\ell+1)^2}\,,\nonumber
 \\\alpha_9&=\frac{4(2\ell^6+23\ell^5+43\ell^4-11\ell^3-45\ell^2+6\ell+6)}{3\ell(\ell+1)^2(2\ell+1)^3}\,,\nonumber
 \\\alpha_{10}&=\frac{4(2\ell^6-11\ell^5-42\ell^4-7\ell^3+46\ell^2+6\ell-12)}{3\ell^2(\ell+1)(2\ell+1)^3}\,,\nonumber
 \\\alpha_{11}&=\frac{2(\ell^4+2\ell^3-5\ell^2-6\ell+4)}{\ell(\ell+1)(2\ell+1)^2}
\end{align}
\begin{align}
 \beta_1&=\frac{4\ell^2(4\ell^3+14\ell^2+14\ell+3)}{3(\ell+1)^3(2\ell+1)^2}\,,\quad \beta_2=\frac{4(4\ell^2+4\ell-3)}{3(2\ell+1)^2}\,,\nonumber
 \\\beta_3&=\frac{4(8\ell^6+36\ell^5+46\ell^4+5\ell^3-9\ell^2+7\ell-3)}{3(\ell+1)^3(2\ell+1)^3}\,,\nonumber 
 \\\beta_4&=\frac{4(2\ell^4+9\ell^3+15\ell^2+7\ell-3)}{3(\ell+1)^3(2\ell+1)^3}\,,\quad \beta_5=\frac{4(-4\ell^4-8\ell^3+7\ell^2+11\ell-6)}{3(\ell+1)^2(2\ell+1)^2}\,,\nonumber
 \\\beta_6&=\frac{4(\ell-1)^2(2\ell^2+7\ell+6)}{3(\ell+1)^3(2\ell+1)^2}\,,\quad \beta_7=\frac{8(8\ell^4+22\ell^3+7\ell^2-10\ell+3)}{3(\ell+1)^2(2\ell+1)^3}\nonumber
 \\\beta_8&=\frac{8\ell(4\ell^3+12\ell^2+7\ell-3)}{(\ell+1)^2(2\ell+1)^3}\,,\quad \beta_9=\beta_{10}=\frac{8\ell(2\ell+3)}{(\ell+1)(2\ell+1)^3}\,,
\end{align}
\begin{align}
 \gamma_1&=\frac{4(\ell+1)^2(4\ell^3-2\ell^2-2\ell+1)}{3\ell^3(2\ell+1)^2}\,,\quad \gamma_2=\frac{2(-4\ell^4+2\ell^3-8\ell+4)}{3\ell^3(2\ell+1)^3}\,,\nonumber
 \\\gamma_3&=\frac{4(8\ell^6+12\ell^5-14\ell^4-21\ell^3+12\ell^2+12\ell-6)}{3\ell^3(2\ell+1)^3}\,,\nonumber
 \\\gamma_4&=\frac{4(\ell+2)^2(2\ell^2-3\ell+1)}{3\ell^3(2\ell+1)^2}\,,\quad \gamma_5=\frac{4(4\ell^4+8\ell^3-7\ell^2-11\ell+6)}{3\ell^2(2\ell+1)^2}\,,\nonumber
 \\\gamma_6&=\frac{8(8\ell^4+10\ell^3-11\ell^2-10\ell+6)}{3\ell^2(2\ell+1)^3}\,,\quad \gamma_7=\gamma_9=\frac{8(\ell+1)(2\ell-1)}{\ell(2\ell+1)^3}\,,\nonumber
 \\\gamma_8&=\frac{8(4\ell^4+4\ell^3-5\ell^2-3\ell+2)}{\ell^2(2\ell+1)^3}\,,                                                                                                                     
\end{align}
\begin{align}
 \delta_1&=\frac{2(-3\ell^8-12\ell^7-12\ell^6+6\ell^5+13\ell^4+2\ell^3+2\ell-2)}{3\ell^3(\ell+1)^3(2\ell+1)^4}\,,\nonumber
 \\\delta_2&=\frac{4(2\ell^8+15\ell^7+23\ell^6-11\ell^5-23\ell^4+\ell^3-11\ell^2+4)}{3\ell^3(\ell+1)^2(2\ell+1)^4}\,,\nonumber
 \\\delta_3&=\frac{4(2\ell^8+\ell^7-26\ell^6-54\ell^5-8\ell^4+64\ell^3+44\ell^2-13\ell-10)}{3\ell^2(\ell+1)^3(2\ell+1)^4}\,,\nonumber
 \\\delta_4&=\frac{8(\ell^6+3\ell^5+5\ell^4+5\ell^3-8\ell^2-10\ell+6)}{3\ell^2(\ell+1)^2(2\ell+1)^4}\,,\quad\delta_5=\frac{8(2\ell^6+7\ell^5+3\ell^4-15\ell^3-15\ell^2+3\ell+3)}{3\ell(\ell+1)^2(2\ell+1)^5}\nonumber
 \\\delta_6&=\frac{8(2\ell^6+5\ell^5-2\ell^4-3\ell^3+8\ell^2+\ell-2)}{3\ell^2(\ell+1)(2\ell+1)^5}\,,\quad \delta_7=\frac{4(3\ell^4+6\ell^3-5\ell^2-8\ell+4)}{\ell(\ell+1)(2\ell+1)^3}\,,\nonumber
 \\\delta_8&=\frac{4(3\ell^4+6\ell^3-\ell^2-4\ell+2)}{\ell(\ell+1)(2\ell+1)^4}\,,\quad \delta_9=\frac{8(\ell^6+3\ell^5+3\ell^4+\ell^3-6\ell^2-6\ell+4)}{\ell^2(\ell+1)^2(2\ell+1)^3}\,,\nonumber
 \\\delta_{10}&=\frac{4(2\ell^4+3\ell^3-5\ell^2-4\ell+4)}{\ell(2\ell+1)^4}\,,\quad \delta_{11}=\frac{4(2\ell^4+5\ell^3-2\ell^2-7\ell+2)}{(\ell+1)(2\ell+1)^4}\,,\nonumber
 \\\delta_{12}&=\frac{8\ell(\ell+1)}{(2\ell+1)^4}\,,\quad \delta_{13}=\frac{8\ell(2\ell^2-5\ell-9)}{3(2\ell+1)^5}\,,\quad \delta_{14}=\frac{8(2\ell^3+11\ell^2+7\ell-2)}{3(2\ell+1)^5}\,,\nonumber
 \\\delta_{15}&=\frac{4(-\ell^2-\ell+2)}{(2\ell+1)^3}\,,\quad \delta_{16}=\frac{2(-\ell^2-\ell-2)}{(2\ell+1)^4}
\end{align}
\begin{align}
R_1&=-\frac{\ell}{2(\ell+1)}(|\varphi|^2-1)-\frac{\ell+1}{2\ell}(|\chi|^2-1)+\frac{1}{\ell^2+\ell}(\chi_3-\varphi_3)-\frac{2\ell^4+4\ell^3-2\ell-1}{2(2\ell+1)^2(\ell^2+\ell)}(\chi_3-\varphi_3)^2\nonumber
\\&-\frac{2\ell^2+2\ell-1}{2\ell+1}\psi+\frac{1}{2\ell+1}(\chi_3-\varphi_3)\psi-\frac{\ell^2+\ell+1}{(2\ell+1)^2}\psi^2\,,
\\R_2&=\frac{\ell}{2\ell^2+3\ell+1}(|\varphi|^2-1)+\frac{2\ell^2+\ell-1}{2(2\ell^2+1)}(|\chi|^2-1)+\frac{\ell^2+2\ell-1}{(2\ell+1)(\ell^2+\ell)}(\chi_3-\frac{\chi_3^2+\varphi_3^2}{2(2\ell+1)})\nonumber
\\&-\frac{2\ell^3+2\ell^2-3\ell+1}{\ell(2\ell+1)}(\varphi_3-\frac{\varphi_3\chi_3}{2\ell+1})-\frac{\ell+1}{2\ell+1}(\psi+\frac{\psi^2}{2\ell+1})-\frac{2\ell^2+3\ell-1}{(2\ell+1)^2}\varphi_3\psi\nonumber
\\&+\frac{\ell+1}{(2\ell+1)^2}\chi_3\psi\,,
\\R_3&=\frac{2\ell^2+3\ell}{2(2\ell^2+3\ell+1)}(|\varphi|^2-1)-\frac{\ell+1}{2\ell^2+\ell}(|\chi|^2-1)+\frac{\ell^2-2}{(2\ell+1)(\ell^2+\ell)}(\varphi_3-\frac{\varphi_3^2+\chi_3^2}{2(2\ell+1)})\nonumber
\\&-\frac{2\ell^3+4\ell^2-\ell-4}{(2\ell+1)(\ell+1)}(\chi_3-\frac{\chi_3\varphi_3}{2\ell+1})-\frac{\ell}{2\ell+1}(\psi+\frac{\psi^2}{2\ell+1})-\frac{\ell}{(2\ell+1)^2}\varphi_3\psi\nonumber
\\&-\frac{2\ell^2+\ell-2}{(2\ell+1)^2}\chi_3\psi\,,
\end{align}

Equations of motion that follow from the variations of the action (\ref{actionii}) are
\begin{align}
 &\big(1-\frac{1}{\ell^2}+\frac{1}{\ell^2}(\zeta^2+\eta^2)\big)(\zeta^{\prime\prime}+\frac{\zeta^\prime}{r})-\bigg(-\eta^2\big(1+\frac{3}{4\ell^2}+\frac{(M+a_\theta^2)^2}{2\ell^2r^2}-\frac{(N+a_\theta^1)^2}{\ell^2r^2}\big)+\frac{3}{\ell^2}\zeta^2\eta^2\nonumber
 \\&-\frac{7}{4\ell^2}\eta^4+(1-\frac{1}{\ell^2})\frac{(N+a_\theta^1)^2}{r^2}-\frac{1}{\ell^2}{\zeta^\prime}^2-\frac{1}{2\ell^2}{\eta^\prime}^2-(1-\frac{1}{\ell}+\frac{1}{2\ell^2})+(2-\frac{1}{\ell})\zeta^2\bigg)\zeta=0\,,\nonumber
 \\&\big(1-\frac{1}{\ell^2}+\frac{1}{\ell^2}(\zeta^2+\eta^2)\big)(\eta^{\prime\prime}+\frac{\eta^\prime}{r})-\bigg(-\zeta^2\big(1+\frac{3}{4\ell^2}+\frac{(N+a_\theta^1)^2}{2\ell^2r^2}-\frac{(M+a_\theta^2)^2}{\ell^2r^2}\big)+\frac{3}{\ell^2}\zeta^2\eta^2\nonumber
 \\&-\frac{7}{4\ell^2}\zeta^4+(1-\frac{1}{\ell^2})\frac{(M+a_\theta^2)^2}{r^2}-\frac{1}{\ell^2}{\eta^\prime}^2-\frac{1}{2\ell^2}{\zeta^\prime}^2-(1+\frac{1}{\ell}-\frac{3}{2\ell^2})+(2-\frac{1}{\ell}-\frac{2}{\ell^2})\eta^2\bigg)\eta=0\,,\nonumber
 \\&{a^1_\theta}^{\prime\prime}-\frac{{a^1_\theta}^\prime}{r}+(2-\frac{1}{\ell^2})(M+a_\theta^2)\eta^2-(4-\frac{2}{\ell}+\frac{1}{\ell^2})(N+a_\theta^1)\zeta^2=0\,,\nonumber
 \\&{a^2_\theta}^{\prime\prime}-\frac{{a^2_\theta}^\prime}{r}+(2-\frac{1}{\ell^2})(N+a_\theta^1
 )\zeta^2-(4+\frac{2}{\ell}-\frac{1}{\ell^2})(M+a_\theta^2)\eta^2=0\,,
 \label{larger}
\end{align}

\end{document}